\documentclass[1p, final]{elsarticle}

\usepackage{subfig}
\usepackage{amsmath,bm}
\usepackage{color}
\usepackage{enumerate}
\usepackage{multirow}
\usepackage{amssymb}
\usepackage{epstopdf}
\usepackage{epsfig}
\usepackage[table,xcdraw]{xcolor}
\usepackage{algpseudocode,algorithm,algorithmicx}

\usepackage{lineno,hyperref}
\modulolinenumbers[5]

\journal{Computer Methods in Applied Mechanics and Engineering}

\begin{document}

\begin{frontmatter}

\title{Nonlinear Acceleration of Sequential Fully Implicit (SFI) Method for Coupled Flow and Transport in Porous Media}

\author{Jiamin Jiang}
\cortext[mycorrespondingauthor]{Corresponding author}
\ead{jiamin66@stanford.edu}

\author{Hamdi A. Tchelepi}

\address{Department of Energy Resources Engineering, Stanford University}
\address{367 Panama St., Green Earth Sciences 050, Stanford, CA 94305-4007, USA}

\date{October 4, 2018}

\begin{abstract}

The sequential fully implicit (SFI) method was introduced along with the development of the multiscale finite volume (MSFV) framework, and has received considerable attention in recent years. Each time step for SFI consists of an outer loop to solve the coupled system, in which there is one inner Newton loop to implicitly solve the pressure equation and another loop to implicitly solve the transport equations. Limited research has been conducted that deals with the outer coupling level to investigate the convergence performance. In this paper we extend the basic SFI method with several nonlinear acceleration techniques for improving the outer-loop convergence. Specifically, we consider numerical relaxation, quasi-Newton (QN) and Anderson acceleration (AA) methods. The acceleration techniques are adapted and studied for the first time within the context of SFI for coupled flow and transport in porous media. We reveal that the iterative form of SFI is equivalent to a nonlinear block Gauss-Seidel (BGS) process.

The effectiveness of the acceleration techniques is demonstrated using several challenging examples. The results show that the basic SFI method is quite inefficient, suffering from slow convergence or even convergence failure. In order to better understand the behaviors of SFI, we carry out detailed analysis on the coupling mechanisms between the sub-problems. Compared with the basic SFI method, superior convergence performance is achieved by the acceleration techniques, which can resolve the convergence difficulties associated with various types of coupling effects. We show across a wide range of flow conditions that the acceleration techniques can stabilize the iterative process, and largely reduce the outer iteration count.

\end{abstract}

\end{frontmatter}

\section{Introduction}

Numerical reservoir simulation is an essential tool for improving our understanding of underground resources, including oil and gas recovery, groundwater remediation, and $\mathrm{CO_2}$ subsurface sequestration processes. Predicting evolution of the subsurface fluid dynamics requires solving the partial differential equations (PDEs) that represent multiphase flow and transport in natural porous media. The components of these PDEs are tightly coupled and highly nonlinear. In particular, the transport equations with combined viscous and gravitational forces are characterized by nonconvex and nonmonotonic flux functions (Li and Tchelepi 2015). In addition, detailed geological models with heterogeneous coefficients are usually constructed as input to reservoir simulators. These properties make the development of robust, efficient, and accurate discretization and solution schemes quite challenging (Lee et al. 2015; Jiang and Younis 2017).

Several temporal discretization methods are available to solve the mass conservation equations that describe coupled flow and transport (Aziz and Settari 1979). The use of explicit temporal schemes poses severe restrictions on the time-step size, and is usually considered impractical for large heterogeneous three-dimensional problems, in which the Courant-Friedrichs-Lewy (CFL) numbers can vary by several orders of magnitude throughout the domain (Jenny et al. 2009). Therefore, implicit schemes such as the fully implicit (FI) and sequential implicit (SI) methods are preferred in practice. The resulting nonlinear system is usually cast in residual form and solved using a Newton-based solver. For a target time-step, a sequence of nonlinear (Newton) iterations is performed until the solution (within a specified tolerance) is achieved (Wang and Tchelepi 2013). Each iteration involves construction of the Jacobian matrix and solution of the corresponding linear algebraic equations. 


Compared with the FI formulation, sequential type of methods handles flow and transport separately and differently (Li et al. 2003; Lu 2008; Kozlova et al. 2016). This adds flexibility in the choice of discretization scheme, solution strategy and time-stepping for each sub-problem. Specialized solvers that are well suited to the specific characteristics of the governing equations can be employed to achieve optimal performances (Jenny et al. 2006). There are three common sequential methods developed in the reservoir simulation literature. A classical method is based on the implicit-pressure/explicit-saturation (IMPES) scheme (Aziz and Settari 1979; Coats 2000). The pressure equation is solved implicitly to obtain pressure and total flux, while the transport equations are treated explicitly to update saturations. The stability limits in IMPES impose severe restrictions on the time-step size for heterogeneous reservoir models. To improve stability, sequential implicit (SI) methods that also treat the transport equations implicitly were proposed (Watts 1986). The SI methods still exhibit conditional stability due to the semi-implicit treatment of the velocity. In addition, material balance errors of SI may accumulate with time (Moncorgé et al. 2017). To resolve these issues, Jenny et al. (2006) proposed the sequential fully implicit (SFI) method, which iterates on implicitly solving the pressure and transport equations before advancing to the next time-step. In the SFI algorithm, an outer loop is added on top of the two inner loops executed in sequence. The method provides the converged solution that is identical to the solution computed by a FI discretization; thus, it is consistent and mass conservative (Lee et al. 2008).

The SFI method was introduced along with the development of the multiscale finite volume (MSFV) framework (Jenny et al. 2006). Multiscale type of methods is widely employed as a computationally inexpensive alternative to traditional fine-scale solvers for computing conservative approximations of the pressure and velocity fields on highly detailed reservoir models (Jenny et al. 2003; Jenny et al. 2006; Lunati and Jenny 2006; Efendiev et al. 2006; Efendiev et al. 2013; M{\o}yner and Lie 2014). The MSFV framework largely relies on the sequential solutions (SI or SFI) to treat the flow-transport coupling (Lee et al. 2008; Krogstad et al. 2009; Cusini et al. 2015). It should be noted that MSFV can be used as part of a constrained-pressure-residual (CPR) preconditioner to achieve a fully implicit discretization (Cusini et al. 2015). However, it was illustrated that this formulation is not very efficient in practice. The potential performance of MSFV will be fully exploited, only through a sequential solution procedure (M{\o}yner and Lie 2016; Kozlova et al. 2016; Lie et al. 2017).

For many cases of immiscible coupled flow and transport, SFI performs quite well compared with the FI method (Jenny et al. 2006; Lee et al. 2008). On the other hand, if the coupling terms between the flow and transport equations are too strong, the SFI method suffers from convergence difficulties, resulting in large numbers of outer iterations (Cusini et al. 2015; Watanabe et al. 2016; Moncorgé et al. 2017). Worst of all, when convergence failures are encountered, a common remedy is to restart the nonlinear solver with a smaller time-step size (Younis et al. 2010). This procedure often leads to time-step sizes that are very conservative resulting in unnecessarily long computational time and wasted computations (Wang and Tchelepi 2013). Therefore, the development of a reliable and efficient sequential algorithm is critical to achieving superior convergence performance and computational speedup of the MSFV framework.

Previous works in the reservoir simulation literature focus on the manipulations, formulations, spatial/temporal discretizations and solution order of the flow and transport sub-problems in the sequential type of methods. In particular, a number of heuristic and physics-based damping (safeguarding) strategies have been devised to enlarge the convergence radius of the transport sub-problem (Jenny et al. 2009; Younis et al. 2010; Wang and Tchelepi 2013). However, limited research has been conducted that deals with the outer coupling level to investigate the nonlinear convergence. Note that Jeannin et al. (2007) implemented and studied two acceleration techniques for coupled flow and poromechanics simulations. 

The main objective of this paper is to extend the SFI framework with several nonlinear acceleration techniques for improving both the convergence performance and algorithm stability. Specifically, we consider numerical relaxation, quasi-Newton (QN) and Anderson acceleration (AA) methods, which have been intensively exploited in fluid-structure interaction analysis. Quasi-Newton (QN) methods approximate the Jacobian of the SFI system through the generated nonlinear vector sequence. The nonlinear acceleration techniques are adapted and studied for the first time within the context of SFI for coupled flow and transport in porous media. We show that the iterative form of SFI is equivalent to a nonlinear block Gauss-Seidel (BGS) process. As in SFI, BGS proceeds with alternating solution of the sub-problems, subject to coupling transmission conditions (e.g. fixed total flux). We present detailed analysis for the impacts of the coupling degrees between the sub-problems on the outer loop convergence of SFI.

We evaluate the efficiency of the nonlinear acceleration techniques applied to the SFI framework using several complex examples. We focus on heterogeneous problems with immiscible multiphase flow and transport. The numerical experiments show that the basic SFI method is quite inefficient, suffering from slow convergence or even convergence failure. The performance of SFI deteriorates significantly for the cases with high coupling degrees, especially in the presence of gravitational force. The challenge lies in the fact that phase upstream direction may keep switching (flow reversal) between sequential updates of the pressure and saturation fields. This type of discontinuous behavior has large impacts on the outer-loop level, leading to oscillations or divergence of the iterative process. Compared with the basic SFI method, superior convergence performance is achieved by the three acceleration techniques, which are effective in resolving the convergence difficulties associated with strong coupling effects and the flow reversal issue. We show across a wide range of flow conditions that the acceleration techniques can stabilize the iterative process, and largely reduce the outer iteration count. Among the studied coupling methods, Aitken relaxation provides the optimal performance in terms of overall efficiency, because of its low cost for extra computations. Some graphical explanations are provided for the essential idea behind Aitken.

\section{Mathematical model and discretization}

\subsection{Mathematical model}

We consider compressible and immiscible flow and transport with $n_p$ fluid phases. Pressure-dependent functions are incorporated to relate fluid volumes at reservoir and surface conditions. The conservation equation for phase $l$ is
\begin{equation} 
\label{eq:mass_con}
\frac{\partial \left ( \phi b_{l} s_{l} \right )}{\partial t } + \nabla \cdot \left (b_{l} \textbf{u}_{l} \right ) = b_{l} q_{l} 
\end{equation}
where $l \in \left \{ 1,...,n_p \right \}$. $\phi$ is the rock porosity and $t$ is time. $b_{l}$ is the inverse of the phase formation volume factor (FVF). $s_{l}$ is the phase saturation, with the constraint that the sum of saturations is equal to one $\sum_{l} s_{l} = 1 $. $q_{l}$ is the well flow rate (source and sink terms). Without loss of generality, hereafter we ignore $q_{l}$ because it is zero everywhere except at the cell with well. $\textbf{u}_l$ is the phase velocity, which is written as a function of the phase potential gradient $\nabla \Phi_l $ using the extended Darcy's law
\begin{equation} 
\label{eq:phase_vel}
\textbf{u}_l = -k \lambda_l \nabla \Phi_l = -k\lambda_l\left ( \nabla p - \rho_l g \nabla h \right )
\end{equation}
where $k$ is the rock permeability. $p$ is the pressure. Capillary forces are assumed to be negligible so that there is only a single pressure. $g$ is the gravitational acceleration and $h$ is the height. The phase mobility $\lambda_{l} = k_{rl}/\mu_l$. $k_{rl}$ and $\mu_l$ are the relative permeability and viscosity, respectively. The phase density is evaluated through $\rho_l = b_l \rho_l^S $ and $\rho_l^S$ is the surface density.

The phase velocity can also be expressed using a fractional-flow formulation, which involves the total velocity $\textbf{u}_T $, defined as the sum of the phase velocities
\begin{equation} 
\label{eq:tol_vel}
\textbf{u}_T = \sum_l \textbf{u}_l = -k \lambda_T \nabla p + k \sum_l \lambda_l \rho_l g \nabla h 
\end{equation}
Eq. (\ref{eq:tol_vel}) is used to express the pressure gradient as a function of $\textbf{u}_T$ in order to eliminate the pressure variable from Eq. (\ref{eq:phase_vel}), obtaining
\begin{equation} 
\mathbf{u}_l = \frac{\lambda_{l}}{\lambda_T} \mathbf{u}_T + k g \nabla h \sum_{m} \frac{\lambda_{m}\lambda_{l}}{\lambda_T}  \left ( \rho_{l}-\rho_{m} \right)
\end{equation}

\subsection{Discretization for coupled flow and transport}

The coupled multiphase system describes the interplay between viscous and gravitational forces. The system has an intricate mixture of elliptic and hyperbolic characteristics, which explains why this problem is difficult to solve efficiently and accurately (M{\o}yner and Lie 2016). The method of choice for the time discretization is often a backward, first-order Euler scheme. A standard finite-volume scheme (Cao 2002) is employed as the spatial discretization for the conservation equations. A two-point flux approximation (TFPA) is used to approximate the flux at a cell interface. The fully-implicit discretization of a cell is
\begin{equation} 
\label{eq:dis_mass}
\left ( \phi_i b_{l,i} s_{l,i} \right )^{n+1} - \left ( \phi_i b_{l,i} s_{l,i} \right )^{n} + \frac{\Delta t}{V_i} \sum_{j\in adj(i)} \left ( b_{l,ij} F_{l,{ij}} \right )^{n+1} = 0 
\end{equation}
where subscript $i$ denotes quantities associated with cell $i$ and subscript $ij$ denotes quantities associated with the interface between cells $i$ and $j$. $adj(i)$ is the set of neighbors of cell $i$. Superscripts denote the time level. $\Delta t$ is the time-step size. $V_i$ is the volume of cell $i$. $F_{l,ij}^{n+1}$ is constructed as the approximation for the integral of the phase velocity. 

It is often advantageous to reformulate Eq. (\ref{eq:dis_mass}) into one elliptic or parabolic equation for the pressure and several hyperbolic transport equations for the saturations. This reformulation allows to employ discretization approaches which are especially developed and suited for the corresponding type of equation (Helmig et al. 2010). Furthermore, a sequential solution procedure can be applied to reduce the amount of unknowns in each solution step. Both the pressure and transport equations are nonlinear and need to be solved iteratively.

To derive a discrete pressure equation, we first multiply Eq. (\ref{eq:dis_mass}) by $\alpha_l = 1/b_{l}^{n+1} $. Then with the saturation constraint, the summation of the resulting equations gives 
\begin{equation} 
\label{eq:press_eq}
\phi_i^{n+1} - \phi_i^{n} \sum_{l} \left ( \frac{b_{l,i}^{n}}{b_{l,i}^{n+1}}s_{l,i}^{n} \right ) + \frac{\Delta t}{V_i} \sum_{j\in adj(i)} \sum_{l} \frac{b_{l,ij}^{n+1}}{b_{l,i}^{n+1}} F_{l,{ij}}^{n+1} = 0 
\end{equation}
where the saturation dependency at the current time level, $(n+1)$, is eliminated in the accumulation term. We use Eq. (\ref{eq:press_eq}) to replace one of the phase conservation equations. The overall system under the fractional-flow formulation will consist of the pressure equation and $(n_p-1)$ transport equations. Correspondingly, the primary variables are the pressure and $(n_p-1)$ phase saturations. If we neglect the compressibility, Eq. (\ref{eq:press_eq}) can be reduced to the following elliptic form 
\begin{equation} 
\sum_{j\in adj(i)} u_{T,ij}^{n+1} = 0 
\end{equation}
where $u_{T,ij}$ denotes the total flux.

\section{Upwinding schemes for the finite-volume scheme}

In reservoir simulation, first-order schemes are commonly employed to evaluate the phase mobilities for the numerical flux of the finite-volume method. In this work we focus on two upwinding schemes, namely, Phase-Potential Upwinding (PPU) and Implicit Hybrid Upwinding (IHU). 

\subsection{Phase-Potential Upwinding (PPU)}

In PPU, the upstream direction of a fluid phase is determined based on the sign of the discretized phase potential at cell interface. Specifically, the PPU flux reads
\begin{equation} 
F_{l,ij}^{PPU} = T_{ij} \lambda_{l,ij}^{PPU} \Delta \Phi_{l,ij}
\end{equation}
where the total transmissibility $T_{ij}$ that combines the two half-transmissibilities in a harmonic average is
\begin{equation}
T_{ij} = \frac{T_i T_j}{T_i + T_j}
\end{equation}
The two-point half-transmissibility for a general grid is obtained by imposing flux and pressure continuity at the center of the interface. The mobility is evaluated according to the upwinding criterion
\begin{equation} 
\label{eq:PPU}
\lambda_{l,ij}^{PPU} = \left\{ {\begin{array}{*{20}c}
\lambda_{l}(s_{i}),   & \Delta \Phi_{l,ij} > 0 \\ 
\lambda_{l}(s_{j}), &  \mathrm{otherwise}
\end{array}} \right.
\end{equation}
where $s_i = \left \{ s_{k,i} \right \}_{k \in \left \{ 1,...,n_p \right \}}$ refers to the saturations in cell $i$. $\Delta \Phi_{l,ij} = \Delta p_{ij} - g_{l,ij}$ is the phase potential difference with the discrete weights $g_{l,ij} = \rho_l g \Delta h_{ij} $. The PPU flux can also be written as a function of the total flux at the interface 
\begin{equation} 
\label{eq:Fl_tol_v}
F_{l,ij}^{PPU} = \frac{\lambda_{l,ij}^{PPU}}{\lambda_{T,ij}^{PPU}} u_{T,ij} + T_{ij} \sum_{m} \frac{\lambda_{l,ij}^{PPU} \lambda_{m,ij}^{PPU}}{\lambda_{T,ij}^{PPU}} \left ( g_{m,ij} - g_{l,ij} \right )
\end{equation}
where $\lambda_{T,ij}^{PPU} = \sum_{m} \lambda_{m,ij}^{PPU}$. The total velocity discretization is given by
\begin{equation} 
\label{eq:disc_tol_v}
u_{T,ij} = \sum_{m} T_{ij} \lambda_{m,ij}^{PPU} \Delta \Phi_{m,ij} = T_{ij} \lambda_{T,ij}^{PPU} \Delta p_{ij} - T_{ij} \sum_{m} \lambda_{m,ij}^{PPU} g_{m,ij} 
\end{equation}
The resulting total flux is an increasing, but non-differentiable function of $\Delta p_{ij}$. If the total flux is fixed, flow reversal from co-current to counter-current at a cell interface depends on the saturation change in the upwind cell during the transport iterations.

\subsection{Implicit Hybrid Upwinding (IHU)}

As illustrated in Li and Tchelepi (2015) and Lee et al. (2015), PPU could be a source of nonlinear convergence difficulties in the presence of buoyancy. In PPU, small changes in pressure and saturations between two nonlinear iterations may change the sign of the phase potential difference, which results in the switching from co-current to counter-current flow and vice versa. This `flip-flopping' behavior has a significant impact on nonlinear convergence rate (Hamon et al. 2016). Recently, Implicit Hybrid Upwinding (IHU) scheme is proposed and extended for obtaining a smooth numerical flux (Lee et al. 2015; Lee and Efendiev 2016; Hamon and Tchelepi 2016; Hamon et al. 2016). In IHU, the viscous and gravity parts are treated separately. The upwinding of the viscous term is based on the sign of the total velocity, while the directionality in the gravity term is based on density differences. This strategy can lead to a larger convergence radius of the Newton solver and overall faster convergence for two- and three-phase transport problems. The IHU flux is expressed as
\begin{equation} 
F_{l,ij}^{HU} = V_{l,ij} + G_{l,ij}
\end{equation}
where the viscous part $V_{l,ij}$ is defined as
\begin{equation} 
V_{l,ij} = \frac{\lambda_{l,ij}^{V}}{\lambda_{T,ij}^{V}} u_{T,ij} \qquad \textrm{with} \ \ \lambda_{T,ij}^{V} = \sum_{k} \lambda_{k,ij}^{V}
\end{equation}
Note that the discretized total velocity is still evaluated through Eq. (\ref{eq:disc_tol_v}), where PPU is used for mobility upwinding. The mobility $\lambda_{l,ij}^{V}$ is
\begin{equation} 
\label{eq:V_HU}
\lambda_{l,ij}^{V} = \left\{ {\begin{array}{*{20}c}
\lambda_{l}(s_{i}),   & u_{T,ij} > 0 , \\ 
\lambda_{l}(s_{j}), &  \mathrm{otherwise}.
\end{array}} \right.
\end{equation}
The gravity part is written as
\begin{equation} 
\label{eq:G_HU}
G_{l,ij} = T_{ij} \sum_{m} \frac{\left ( \lambda_l \lambda_m \right )_{ij}^G}{\lambda_{T,ij}^G} \left ( g_{m,ij} - g_{l,ij} \right )
\end{equation}
In Eq. (\ref{eq:G_HU}), for a given pair $(l,m)$, the evaluation of the product $\left ( \lambda_l \lambda_m \right )_{ij}^G$ in the mobility ratio is determined by the fluid weights. The evaluation is based on the fact that the heavier fluid goes down and that the lighter fluid goes up (Lee et al. 2015). Specifically, using the ordering of phase densities 
\begin{equation} 
m \leq k \Longrightarrow \rho_m \leq \rho_k 
\end{equation}
we obtain
\begin{equation} 
\left ( \lambda_l \lambda_m \right )_{ij}^G = \left\{ {\begin{array}{*{20}c}
\lambda_{l}(s_{i}) \lambda_{m}(s_{j}),   & (l-m)(h_j-h_i)g >0 \\ 
\lambda_{l}(s_{j}) \lambda_{m}(s_{i}), &  \mathrm{otherwise}
\end{array}} \right.
\end{equation}

To guarantee monotonicity of the numerical flux with respect to its own saturation, the total mobility in the gravity part is given by (Hamon and Tchelepi 2016)
\begin{equation} 
\lambda_{T,ij}^G = \sum_{k} \overline{\lambda}_{k,ij}
\end{equation}
where the phase mobility is a weighted average of the mobilities in cells $i$ and $j$ 
\begin{equation} 
\overline{\lambda}_{k,ij} = \left\{ {\begin{array}{*{20}c} \alpha_{k,ij} \lambda_{k}(s_{i}) + \left ( 1 - \alpha_{k,ij} \right ) \lambda_{k}(s_{j}),  & (h_j - h_i) g > 0 \\ \alpha_{k,ij} \lambda_{k}(s_{j}) + \left ( 1 - \alpha_{k,ij} \right ) \lambda_{k}(s_{i}), &  \mathrm{otherwise}
\end{array}} \right. 
\end{equation}
The coefficients $\alpha_{k,ij}$ are defined as a ratio of the density differences
\begin{equation} 
\alpha_{k,ij} = \frac{g_{k,ij} - g_{1,ij}}{g_{n_p,ij} - g_{1,ij}} \in [0,1]
\end{equation}
where $g_{1,ij}$ (respectively, $g_{n_p,ij}$) is the weight of the lightest (respectively, the heaviest) phase. The IHU flux is consistent and monotone with respect to its own saturation when the total velocity is fixed. 

\section{Sequential Fully Implicit strategy}

Sequential Fully Implicit (SFI) method (Jenny et al. 2006) is a popular solution strategy to handle coupled flow and transport in porous media. For each time-step in SFI, there is an outer coupling loop, and there are two inner loops: one for pressure and one for saturation. At each inner loop, linearized equations are solved and convergence is achieved by the Newton method. For each iteration of the outer loop, the computations proceed as follows: compute the pressure field iteratively to a certain tolerance, and update the total flux, then compute the saturation iteratively (Lunati and Jenny 2006). Note that the total flux is always evaluated through Eq. (\ref{eq:disc_tol_v}), where the discrete phase mobilities are given by the PPU scheme. The updated saturation defines a new mobility field for the subsequent pressure problem. These steps can be iterated until convergence of all variables at the current time level (Kozlova et al. 2016). 

For the sequential methods, if only a single iteration of the outer loop is performed and the transport equations are treated explicitly, we obtain the IMPES method. If the transport equations are also solved implicitly, we obtain a sequential implicit method, which is not unconditionally stable, and not exactly conservative for all fluid phases (Watts 1986; Jenny et al. 2006).

The solution procedure of SFI for one time-step is demonstrated in Algorithm \ref{alg:SFI}. $r_p$ and $r_s$ are the residuals of the pressure and transport equations. $\epsilon$, $\epsilon_p$ and $\epsilon_s$ are the residual tolerances of convergence criteria in the outer, pressure and transport loops, respectively. $J = \left [ \frac{\partial r}{\partial u} \right ] $ is the Jacobian matrix. Compared with SFI, the fully implicit (FI) method solves for all the primary variables simultaneously. The sparse linear system arising from each Newton iteration reads 
\begin{equation} 
\begin{bmatrix}
J_{pp} & J_{ps} \\ 
J_{sp} & J_{ss}
\end{bmatrix}
\begin{bmatrix}
\delta p \\ 
\delta s
\end{bmatrix}
= - 
\begin{bmatrix}
r_p \\ 
r_s
\end{bmatrix}
\end{equation}
where the block matrices $J_{ps}$ and $J_{sp}$ represent the cross-coupling coefficients.

\begin{algorithm}
\caption{Sequential fully implicit method} \label{alg:SFI}
\begin{algorithmic}[1]
\State $\nu = 1$, initialize $p^{(\nu)} = p^n$, $s^{(\nu)} = s^n$ 
\While{$\textrm{max}\left ( \left | r_p \right |,\left | r_s \right | \right ) > \epsilon$ }
\Comment{Outer coupling loop}
\smallskip
\State $\nu_p = 1$, $p^{\nu_p} = p^{(\nu)}$ 
\While{$\left | \delta p \right | > \epsilon_p $ } \Comment{Pressure loop}
\State Solve linearized pressure equation:
\State $J_p \delta p = -r_p$
\State $p^{\nu_p + 1} = p^{\nu_p} + \delta p$
\State $\nu_p \leftarrow \nu_p + 1$
\EndWhile 

\medskip
\State Compute total flux by summing phase fluxes.
\medskip

\State $\nu_s = 1$, $s^{\nu_s} = s^{(\nu)}$ 
\While{$\left | \delta s \right | > \epsilon_s $ } \Comment{Saturation loop}
\State Solve linearized transport equations:
\State $J_s \delta s = -r_s$
\State $s^{\nu_s + 1} = s^{\nu_s} + \delta s$
\State $\nu_s \leftarrow \nu_s + 1$
\EndWhile 

\State $p^{(\nu + 1)} = p^{\nu_p }$, $s^{(\nu + 1)} = s^{\nu_s }$
\State $\nu \leftarrow \nu + 1$

\EndWhile 
\end{algorithmic}
\end{algorithm}

\subsection{Block Gauss-Seidel formulation}

From Eqs. (\ref{eq:press_eq}) and (\ref{eq:Fl_tol_v}), we can see that the pressure equation is tightly coupled to the highly nonlinear transport through the mobility terms. In the presence of buoyancy, and sharp saturation fronts propagating in the domain, the coupling between flow and transport can become quite strong. The SFI solution strategy, although simple to implement, can suffer from slow convergence, or even fail to converge. Nonlinear acceleration techniques can be utilized to improve the convergence of SFI. To present the design of implicit coupling algorithms, we first recall the global, coupled problem
\begin{equation} 
\left\{\begin{matrix}
r_p \left ( p,s \right ) = 0 , \\ 
r_s \left ( p,s \right ) = 0 .
\end{matrix}\right.
\end{equation}
It can be shown that the iterative form of SFI is equivalent to the block Gauss-Seidel (BGS) process, namely
\begin{equation} 
\label{eq:n_GS}
\begin{cases}
p^{(\nu+1)} = \mathcal{P} \left ( p^{(\nu)},s^{(\nu)} \right ) , \\ 
s^{(\nu+1)} = \mathcal{T} \left ( p^{(\nu+1)},s^{(\nu)} \right ) .
\end{cases}
\end{equation}
where the operators $\mathcal{P}$ and $\mathcal{T}$ represent respectively the pressure and transport solvers. The transport solver works on the results from the pressure solver. Note that the coupling is also subject to the transmission condition with fixed total flux. The counter $\nu$ denotes an outer iteration, which should not be confused with the inner iterations over the individual solvers. The BGS coupling involves repeated applications of the update formulas (\ref{eq:n_GS}). The global problem is converged when the solution to Eq. (\ref{eq:n_GS}) is consistent and both sub-problems are converged (Jenny et al. 2006). 

The BGS iteration can be written in compact form as 
\begin{equation} 
\label{eq:c_n_GS}
\widetilde{s}^{(\nu+1)} = \mathcal{T} \left ( p^{(\nu+1)} \right ) = \mathcal{T} \Big ( \mathcal{P} \left ( s^{(\nu)} \right ) \Big ) = \mathcal{T} \circ \mathcal{P} \left ( s^{(\nu)} \right ) 
\end{equation}
where a tilde $\widetilde{(\cdot )}$ denotes the current unmodified solution during the iterative process. The nonlinear operator $\mathcal{T} \circ \mathcal{P} $ takes an input vector $s^{(\nu)} $ and generates an output vector $\widetilde{s}^{(\nu+1)} $ of the same size.

To enable a fully implicit scheme, a residual operator is introduced from Eq. (\ref{eq:c_n_GS})
\begin{equation} 
\label{eq:resi_n_GS}
r^{(\nu+1)} = \widetilde{s}^{(\nu+1)} - s^{(\nu)} = \mathcal{T} \circ \mathcal{P} \left ( s^{(\nu)} \right ) - s^{(\nu)} 
\end{equation}
Iterative correction is required to bring the pressure and saturation responses into balance so that the residual (\ref{eq:resi_n_GS}) vanishes.

\section{Nonlinear acceleration techniques}

In this work, we will show that direct application of the SFI solution strategy (repeated steps of BGS) may lead to oscillations and even to a divergent iterative process. We consider several ways of accelerating the convergence of the BGS iterations within SFI and apply such methods to the problem of coupled flow and transport. The objective is to reduce the number of outer coupling iterations. 

\subsection{Aitken relaxation}


Aitken relaxation is widely used in the literature of fluid-structure interaction, to accelerate the convergence of BGS by modifying the current solution before it is passed to the other field in the next iteration. This method has proven to be simple and efficient (Kuttler and Wall 2008; Erbts and DuSter 2012).

To derive a relaxation step, we first express the new modified solution as
\begin{equation} 
\label{eq:1_Aitken}
s^{(\nu+1)} = s^{(\nu)} + \Delta s^{(\nu)} 
\end{equation}
where $\Delta s^{(\nu)}$ is the solution difference between the current and previous iterations. $\Delta s^{(\nu)}$ is then replaced by the current residual $r^{(\nu+1)}$ multiplied by a scalar coefficient
\begin{equation} 
\label{eq:2_Aitken}
\Delta s^{(\nu)} = \omega^{(\nu)} r^{(\nu+1)} = \omega^{(\nu)} \left ( \widetilde{s}^{(\nu+1)} - s^{(\nu)}  \right ) 
\end{equation}
where $\omega^{(\nu)}$ denotes the dynamically varying relaxation factor. Substituting (\ref{eq:2_Aitken}) into (\ref{eq:1_Aitken}), the modified solution becomes
\begin{equation} 
s^{(\nu+1)} = \left ( 1 - \omega^{(\nu)} \right ) s^{(\nu)} + \omega^{(\nu)} \widetilde{s}^{(\nu+1)} 
\end{equation}

A simple and ineffective method is to choose a static value such that $\omega^{(\nu)} = \omega$ for all time steps. The relaxation factor has to be small enough to keep the iteration from diverging, but as large as possible in order to make the best use of the new solution and to avoid unnecessary coupling iterations. The optimal $\omega$ value is problem specific and not known $\textit{a-priori}$. Furthermore, even the optimal static value will lead to more iterations than a suitable dynamic relaxation factor (Kuttler and Wall 2008).

Aitken's $\Delta^2$ method calculates a dynamic factor which is adapted in every iteration based on the previous iterations. The notation $\Delta^2$ is due to the fact that the denominator is a difference of a difference, i.e. $\Delta \left ( \Delta  \right )$. For the scalar case, the formula is given by (Kuttler and Wall 2008) 
\begin{equation} 
\label{eq:Aitken_33_f}
s^{(\nu+2)} = \frac{s^{(\nu)}\widetilde{s}^{(\nu+2)} - \widetilde{s}^{(\nu+1)} s^{(\nu+1)}}{s^{(\nu)} - \widetilde{s}^{(\nu+1)} - s^{(\nu+1)} + \widetilde{s}^{(\nu+2)}} 
\end{equation}
and with 
\begin{equation} 
s^{(\nu+2)} = s^{(\nu+1)} + \omega^{(\nu+1)} \left ( \widetilde{s}^{(\nu+2)} - s^{(\nu+1)} \right ) 
\end{equation}
the relaxation factor is updated as
\begin{equation} 
\omega^{(\nu+1)} = -\omega^{(\nu)} \frac{r^{(\nu+1)}}{r^{(\nu+2)} - r^{(\nu+1)}} 
\end{equation}

It is not possible to divide by $\left ( r^{(\nu+2)} - r^{(\nu+1)} \right )$ in the vector case. Instead Irons and Tuck (1969) suggest the form of 
\begin{equation} 
\label{eq:Aitken_11}
\omega^{(\nu+1)} = -\omega^{(\nu)} \frac{\left ( r^{(\nu+1)} \right )^T \left ( r^{(\nu+2)} - r^{(\nu+1)} \right ) }{\left | r^{(\nu+2)} - r^{(\nu+1)} \right |^2 } 
\end{equation}
This is a projection of the participating vectors in $\left ( r^{(\nu+2)} - r^{(\nu+1)} \right ) $ direction and perform the scalar extrapolation with the projected values. 

Some graphical explanations are provided for the essential idea behind Aitken's $\Delta^2$ method in Appendix A. The method is quite straightforward to implement, needs minimum processor and storage resources (Erbts and DuSter 2012). Two previous steps are required in Eq. (\ref{eq:Aitken_11}) and thus it is not possible to compute $\omega$ after the first cycle. The first iteration is carried out using a constant factor $\omega^{(0)}$. Values of $\omega < 1$ stabilize the iteration while values of $\omega > 1$ accelerate the iteration. In this work we use the range $\omega \in (0,1]$ as a threshold for the dynamic relaxation factor.

\subsection{Quasi-Newton}

The BGS solver can be recast as a root-finding problem (\ref{eq:resi_n_GS}) to be tackled by the Newton method. In order to determine the new increment $\Delta s^{(\nu)}$, a linear equation system has to be solved
\begin{equation} 
J^{(\nu)} \Delta s^{(\nu)} = -r^{(\nu+1)} 
\end{equation}
where $J^{(\nu)}$ denotes the Jacobian matrix of the residual operator.

To achieve a more efficient solution process, quasi-Newton (QN) method is applied to the residual (\ref{eq:resi_n_GS}). The QN method approximates the Jacobian system directly from the generated nonlinear vector sequence (Degroote et al. 2010). The QN update with the approximation of the Jacobian inverse is
\begin{equation} 
s^{(\nu+1)} = s^{(\nu)} + \left ( \widehat{\left. \frac{\mathrm{d} r}{\mathrm{d} s} \right|_{s^{(\nu)}}} \right )^{-1} \left ( -r^{(\nu+1)} \right ) 
\end{equation}
The inverse of the Jacobian does not have to be created explicitly, and can be approximated through solving an unconstrained form of the least-squares problem
\begin{equation} 
\label{eq:unc_ls}
\underset{\gamma=\left ( \gamma_0, ..., \gamma_{m_{\nu}-1} \right )^T}{\textrm{min}} \left \| r^{(\nu+1)} - \Delta R^{(\nu)} \gamma \right \|_2 
\end{equation}
with the solution denoted by $\gamma^{\nu} =\left ( \gamma_0^{\nu},...,\gamma_{m_{\nu}-1}^{\nu} \right )^T $. In the following, the Euclidean norm $\left \| \ \cdot \ \right \|_2 $ on $\mathbb{R}^n$ is used throughout.

Some details of QN for a general fixed-point problem are presented in Appendix B. The QN method is adapted to the SFI strategy for coupled flow and transport. The algorithmic description is given in Algorithm \ref{alg:QN_SFI}. The QN update corresponds to the $\textit{first Broyden update}$ (Fang and Saad 2009). $m$ is the number of previous iterations used to obtain the secant information. If $m$ is small, then the secant information may be too limited to provide desirably fast convergence. However, if $m$ is large, the least-squares problem may be undesirably badly conditioned. Moreover, outdated secant information from previous iterations may be kept, leading to convergence degradation (Walker and Ni 2011). Therefore a proper choice of $m$ is likely to be problem-dependent. Note that in the QN algorithm (\ref{alg:QN_SFI}), input and output vectors from several previous iterations are combined to provide a better prediction for the next iteration.

\begin{algorithm}
\caption{Quasi-Newton for the SFI strategy} \label{alg:QN_SFI}
\begin{algorithmic}[1]
\State Set $\omega^{(0)} $ and $m \geq 1$
\State Initialize $p^{(0)} = p^n$, $s^{(0)} = s^n$ 
\State $\widetilde{s}^{(1)} = \mathcal{T} \circ \mathcal{P} \left ( s^{(0)} \right ) $
\State $r^{(1)} = \widetilde{s}^{(1)} - s^{(0)} $
\State $s^{(1)} = s^{(0)} + \omega^{(0)} r^{(1)} $
\State $\nu = 1$
\While{$\textrm{max}\left ( \left | r_p \right |,\left | r_s \right | \right ) > \epsilon$ } 
\smallskip
\State Set $m_{\nu} = \textrm{min} \left \{ m, \nu \right \}$ 
\State $\widetilde{s}^{(\nu+1)} = \mathcal{T} \circ \mathcal{P} \left ( s^{(\nu)} \right ) $
\State $r^{(\nu+1)} = \widetilde{s}^{(\nu+1)} - s^{(\nu)} $
\State $\Delta S^{(\nu)} = \left ( \Delta s_{\nu-m_{\nu}}, ..., \Delta s_{\nu-1} \right ) $, where $\Delta s_i =s^{(i+1)} - s^{(i)}$ for each $i$. 
\State $\Delta R^{(\nu)} = \left ( \Delta r_{\nu-m_{\nu}+1}, ..., \Delta r_{\nu} \right ) $, where $\Delta r_i = r^{(i+1)} - r^{(i)} $ for each $i$. 
\State $\gamma^{(\nu)} = \left ( \left ( \Delta S^{(\nu)} \right )^T \Delta R^{(\nu)} \right )^{-1} \left ( \Delta S^{(\nu)} \right )^T r^{(\nu+1)} $
\State $s^{(\nu+1)} = s^{(\nu)} + \omega^{(0)} r^{(\nu+1)} - \left ( \Delta S^{(\nu)} + \omega^{(0)} \Delta R^{(\nu)} \right ) \gamma^{(\nu)} $ 
\State $\nu \leftarrow \nu + 1$
\smallskip
\EndWhile 
\end{algorithmic}
\end{algorithm}

\subsection{Anderson acceleration }

The Anderson approach finds the next Newton direction for (\ref{eq:resi_n_GS}) using the optimal combination of the input and output vectors from several previous iterations. Fang and Saad (2009) discussed a remarkable relationship between Anderson acceleration (AA) and certain QN methods. In particular, Anderson acceleration is a special case of the Broyden-like class methods. We separately discuss AA here for its popularity in the computational mechanics literature as a means of improving convergence of fixed-point methods. In each AA iteration, the solution for the least-squares problem with the unconstrained form (\ref{eq:unc_ls}) is based on QR decomposition as suggested in Walker and Ni (2011). The QR decomposition of $\Delta R^{(\nu)}$ is given by
\begin{equation} 
\Delta R^{(\nu)} = \mathcal{Q}^{\nu} \times \mathcal{R}^{\nu} 
\end{equation}
Then the solution can be expressed as
\begin{equation} 
\gamma^{(\nu)} = \textrm{arg} \ \underset{\gamma}{\textrm{min}} \left \| \left ( \mathcal{Q}^{\nu} \right )^T r^{(\nu+1)} - \mathcal{R}^{\nu} \gamma \right \|_2 
\end{equation}
which is obtained by solving an $m_{\nu} \times m_{\nu}$ triangular system $\mathcal{R}^{\nu} \gamma = \left ( \mathcal{Q}^{\nu} \right )^T r^{(\nu+1)} $. 

Only the `economy' (`thin') QR decomposition is necessary. As the algorithm proceeds, the QR decomposition of $\Delta R^{(\nu)}$ can be efficiently achieved through updating that of $\Delta R^{(\nu-1)}$: $\Delta R^{(\nu)}$ is obtained from $\Delta R^{(\nu-1)}$ by appending a new column on the right and possibly dropping one column from the left (An et al. 2017). The details of the AA algorithm can be found in Walker and Ni (2011).

\section{Results}

We validate the effectiveness of the nonlinear acceleration techniques applied to the SFI framework for challenging heterogeneous models with immiscible multiphase flow and transport. The rock is slightly compressible and the oil and water phases are assumed to have constant compressibility so that the inverse FVFs are in the form
\begin{equation} 
b_l = b_{l}(p_0) e^{\left ( p - p_0 \right ) c_l} 
\end{equation}
where $p_0$ is a reference pressure and $c_l$ the compressibility factor for phase $l$.

The relative permeability curves characterize the ability of a fluid phase to flow in the presence of another phase in a specific type of porous media. For a three-phase system, it is common to assume that the relative permeabilities of the gas $k_{rg}(s_g)$ and water $k_{rw}(s_w)$ phases depend on their respective saturations only, while the oil relative permeability depends on both saturations (Hamon and Tchelepi 2016). An interpolation method is usually used to compute $k_{ro}$ from two-phase oil-gas $k_{rog}\left ( s_o \right )$ and oil-water $k_{row}\left ( s_o \right )$ data, see for example the works by Stone (1973) and Baker (1988). As discussed in Lee and Efendiev (2016), the method choice can affect the nonlinear convergence of the underlying discretization scheme.

We employ the weighted interpolation model introduced by Baker (1988)
\begin{equation} 
k_{ro}\left ( s_w,s_o,s_g \right ) = k_{row}\left ( s_o \right ) w_w \left ( s_w,s_g \right ) + k_{rog}\left ( s_o \right ) w_g \left ( s_w,s_g \right )
\end{equation}
and 
\begin{equation} 
w_w \left ( s_w,s_g \right ) + w_g \left ( s_w,s_g \right ) = 1
\end{equation}
which is the default setting of the commercial simulator Eclipse (Schlumberger 2013). This choice ensures positive values and continuous derivatives provided that $k_{row}$, $k_{rog}$ fulfill the same criterion (Baker 1988; Lee and Efendiev 2016).

We consider simple relative permeabilities given by Corey functions for a pair of wetting and non-wetting fluids,
\begin{equation} 
k_{rw}\left ( s_w \right ) = s_{w}^{n}, \quad k_{rg}\left ( s_g \right ) = s_{g}^{n}, \quad k_{row}\left ( s_o \right ) = k_{rog}\left ( s_o \right ) = s_{o}^{n} 
\end{equation}
where $n \in \left \{ 2, 3 \right \}$ in this paper. In this way, the relative permeability of each phase is a function only of its own saturation.

\subsection{Case 1: two-phase, pure gravitational, square model }

We consider the scenario with pure gravitational force. The specification of the base model is shown in Table \ref{tab:specification}. Quadratic relative permeability functions are used. We specify a fixed simulation schedule with uniform time intervals. A simple time-stepping strategy is employed individually for each time interval: if the outer coupling loop of the SFI method fails to converge, time-step is reduced by half until it converges. The simulation control parameters are summarized in Table \ref{tab:control}. The following abbreviations are adopted for all the test cases: BGS for block Gauss-Seidel (Sequential Fully Implicit) without any relaxation or acceleration, Aitken for Aitken's $\Delta^2$ method, QN for quasi-Newton method, and Anderson for Anderson acceleration. In the Aitken method, the relaxation factor for the first iteration is $\omega^{(0)} = 0.5 $. In the QN and Anderson methods, $m = 3 $ and $\omega^{(0)} = 0.8 $. Here we focus on the convergence behavior of the outer loop, and thus the details regarding the Newton solver for achieving the convergence of the sub-problems are neglected.

\begin{table}[!htb]
\centering
\caption{Specification of the base model}
\label{tab:specification}
\begin{tabular}{|c|c|c|}
\hline
Parameter                  &  Value           & Unit      \\ \hline
NX / NZ                    &  60 / 60         &           \\ \hline
LX / LY / LZ               &  600 / 10 / 600  & ft        \\ \hline
Initial water saturation   &  0.001           &           \\ \hline
Initial pressure           &  2000            & psi       \\ \hline
Rock permeability          &  100             & md        \\ \hline
Rock porosity              &  0.1             &           \\ \hline
Water density              &  1000            & $\textrm{kg}/\textrm{m}^3$  \\ \hline
Oil   density              &  500             & $\textrm{kg}/\textrm{m}^3$  \\ \hline
Water viscosity            &  1               & cP        \\ \hline
Oil   viscosity            &  4               & cP        \\ \hline
Water compressibility      &  0               & 1/psi     \\ \hline
Oil   compressibility      &  6.895e-6        & 1/psi     \\ \hline
Rock  compressibility      &  1e-6            & 1/psi     \\ \hline
Reference pressure         &  0               & psi       \\ \hline
\end{tabular}
\end{table}

\begin{table}[!htb]
\centering
\caption{Simulation control parameters}
\label{tab:control}
\begin{tabular}{|c|c|c|}
\hline
Parameter                 & Value          & Unit   \\ \hline
Time-interval size        & 50             & day    \\ \hline
Total simulation time     & 400            & day    \\ \hline
Maximum number of outer iterations      & 30         &     \\ \hline
Convergence tolerance of outer loop     & 0.001      &     \\ \hline
\end{tabular}
\end{table}

\subsubsection{Case 1a: lock-exchange, homogeneous }

\begin{figure}[!htb]
\centering
\includegraphics[scale=0.7]{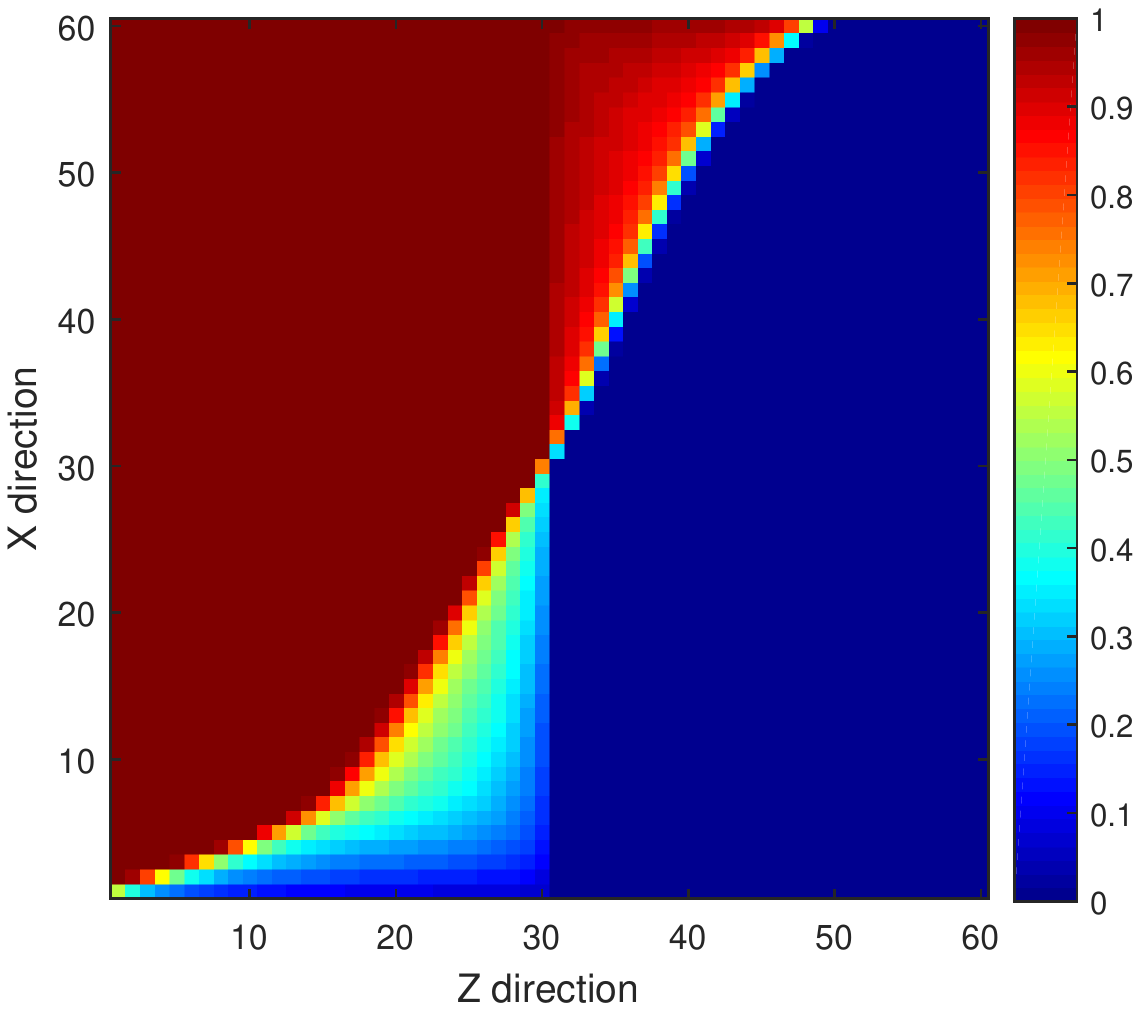}
\caption{Oil saturation profile of Case 1a }
\label{fig:s_h_le_pg}
\end{figure}

\begin{figure}[!htb]
\centering
\includegraphics[scale=0.6]{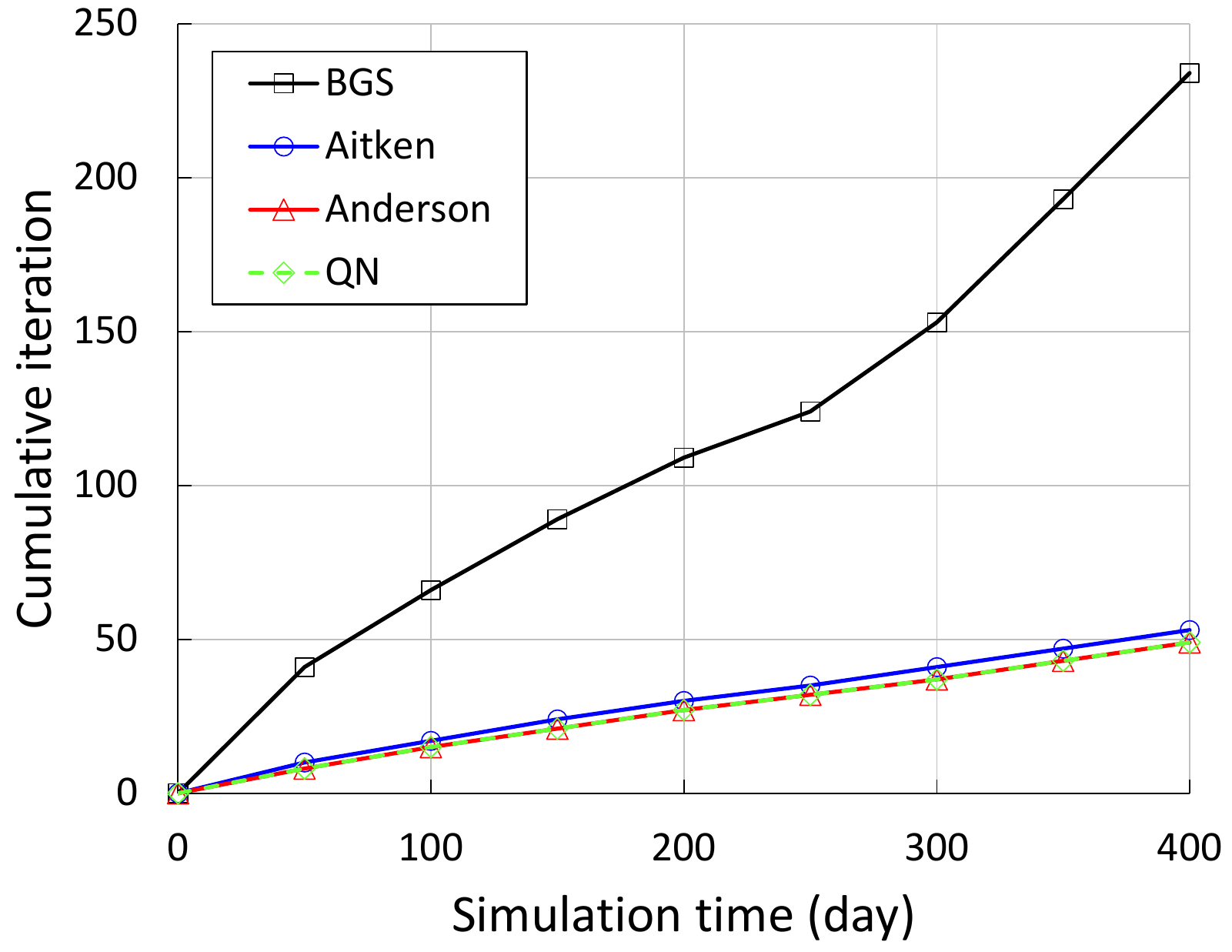}
\caption{Cumulative number of outer iterations versus simulation time of Case 1a }
\label{fig:cN_h_le_pg}
\end{figure}

We first test a lock-exchange problem on the homogeneous square model. The gravity is towards the X direction. Oil initially occupies the left half of the domain, while water fills the right half. The lock-exchange problem is very challenging, because gravity significantly contributes to the total velocity and induces a re-circulation flow pattern (Lunati and Jenny 2008). The oil saturation profile at the end of simulation and the cumulative number of outer iterations versus simulation time for the PPU flux are plotted in \textbf{Fig. \ref{fig:s_h_le_pg}} and \textbf{Fig. \ref{fig:cN_h_le_pg}}, respectively.

From the results we can see that the basic BGS (SFI) process suffers from severe restrictions on the allowable time-step size. Even though the model is homogeneous and has no wells, strong coupling exists between the flow and transport sub-problems. Compared with BGS, the three nonlinear acceleration techniques do not require any time-step cuts, leading to uniform convergence for the time intervals.

\subsubsection{Case 1b: lock-exchange, heterogeneous }

\begin{figure}[!htb]
\centering
\subfloat[Permeability (md)]{
\includegraphics[scale=0.5]{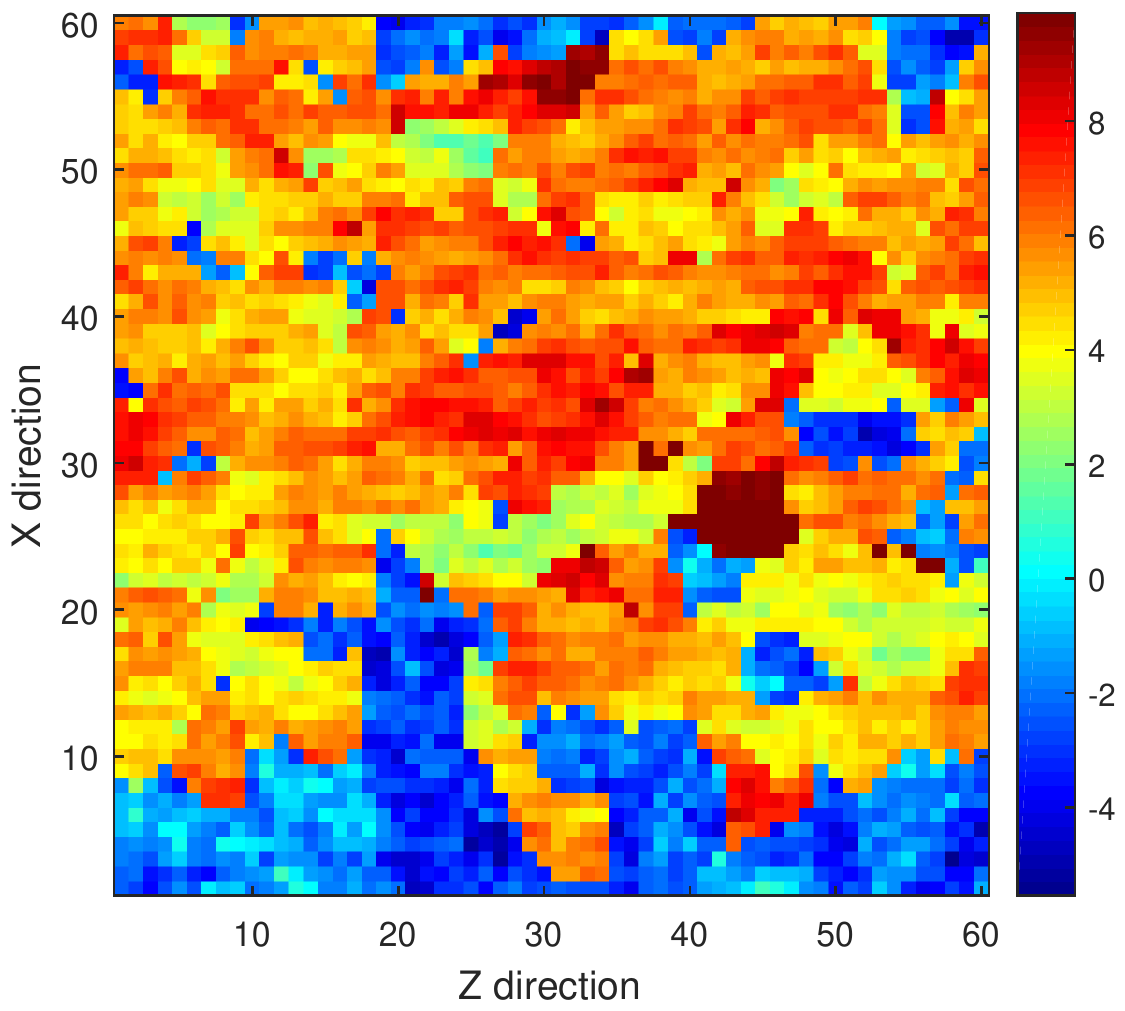}}
\subfloat[Porosity]{
\includegraphics[scale=0.5]{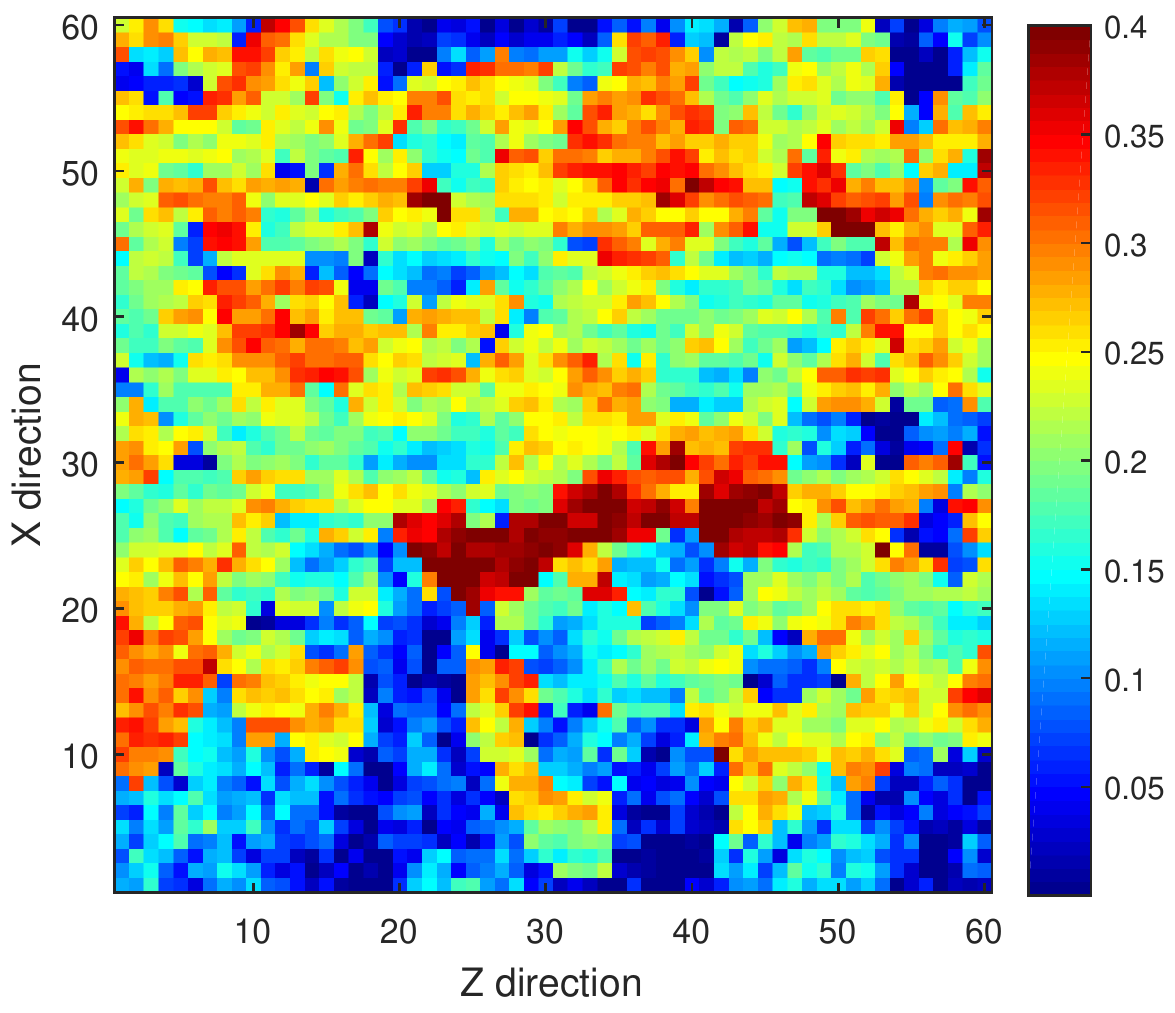}}
\caption{Random permeability (log) and porosity fields for the heterogeneous square model }
\label{fig:perm_poro_sq}
\end{figure}

\begin{figure}[!htb]
\centering
\includegraphics[scale=0.7]{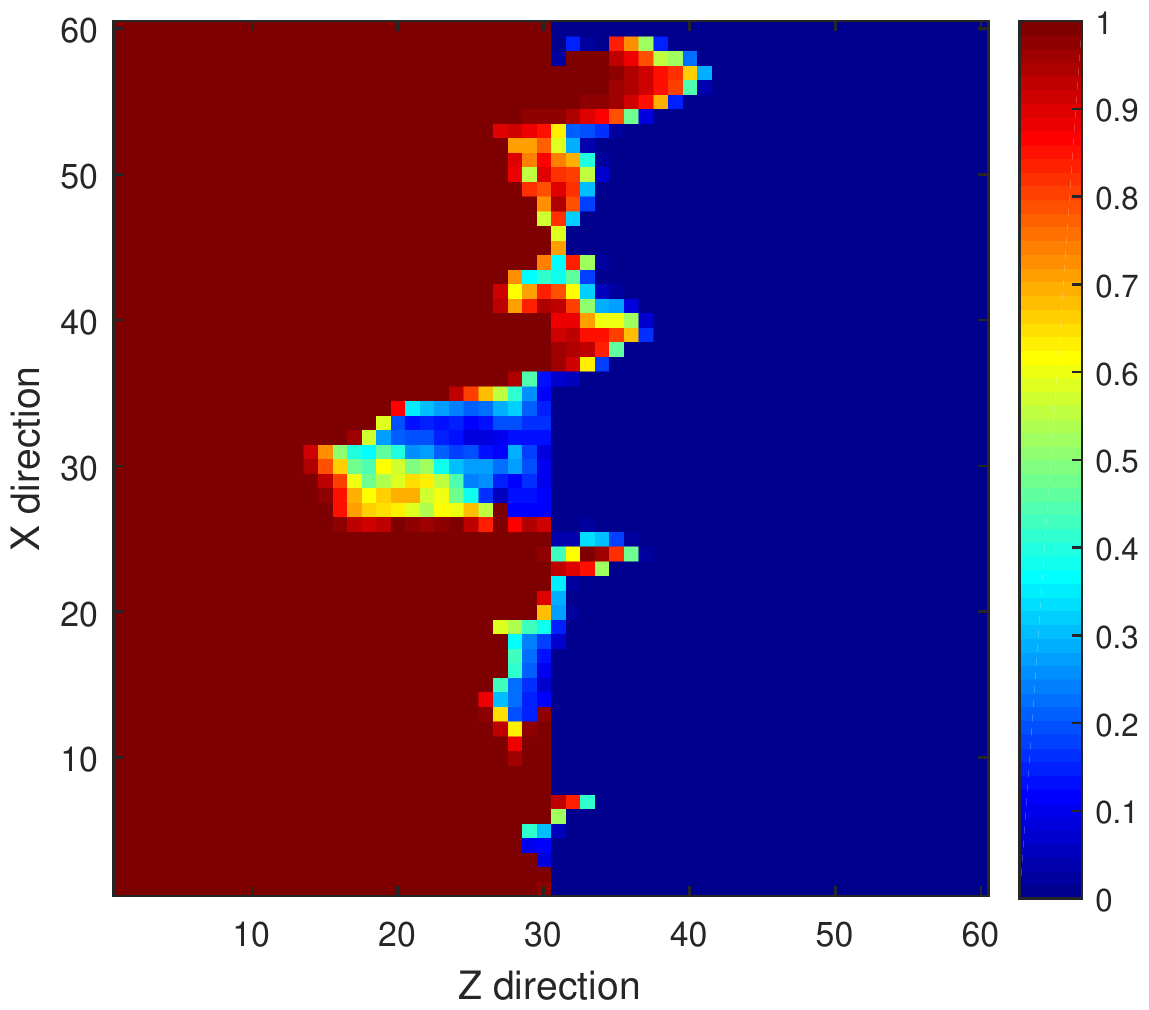}
\caption{Oil saturation profile of Case 1b }
\label{fig:s_he_le_pg}
\end{figure}

\begin{figure}[!htb]
\centering
\includegraphics[scale=0.6]{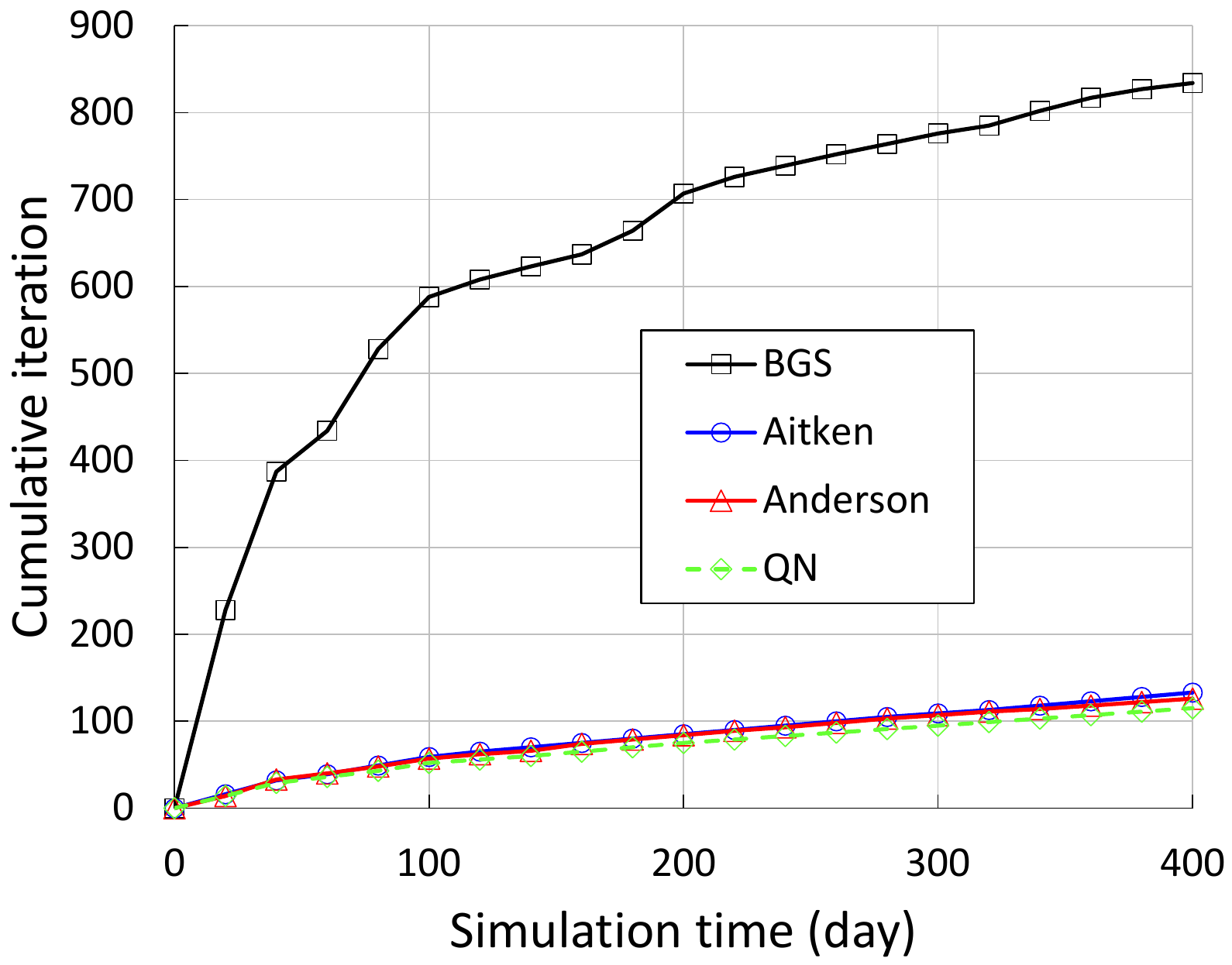}
\caption{Cumulative number of outer iterations versus simulation time of Case 1b }
\label{fig:cN_he_le_pg}
\end{figure}

Oil density is changed to 800 $\textrm{kg}/\textrm{m}^3$ and time-interval size becomes 20 days for the following two-phase cases. The lock-exchange problem is tested on a heterogeneous square model. The random rock properties of the model are shown in \textbf{Fig. \ref{fig:perm_poro_sq}}. The oil saturation profile and the cumulative number of outer iterations versus simulation time for the IHU flux are plotted in \textbf{Fig. \ref{fig:s_he_le_pg}} and \textbf{Fig. \ref{fig:cN_he_le_pg}}, respectively. Compared to Case 1a, the heterogeneous model is much more challenging for BGS to achieve convergence of the outer coupling loop, because of a large variation in the CFL numbers throughout the domain. In particular, BGS initially suffers from severe convergence difficulty, and the situation becomes alleviated afterwards. The other three methods again show a smooth convergence behavior.

\subsubsection{Case 1c: counter-current flow, heterogeneous }

\begin{figure}[!htb]
\centering
\includegraphics[scale=0.7]{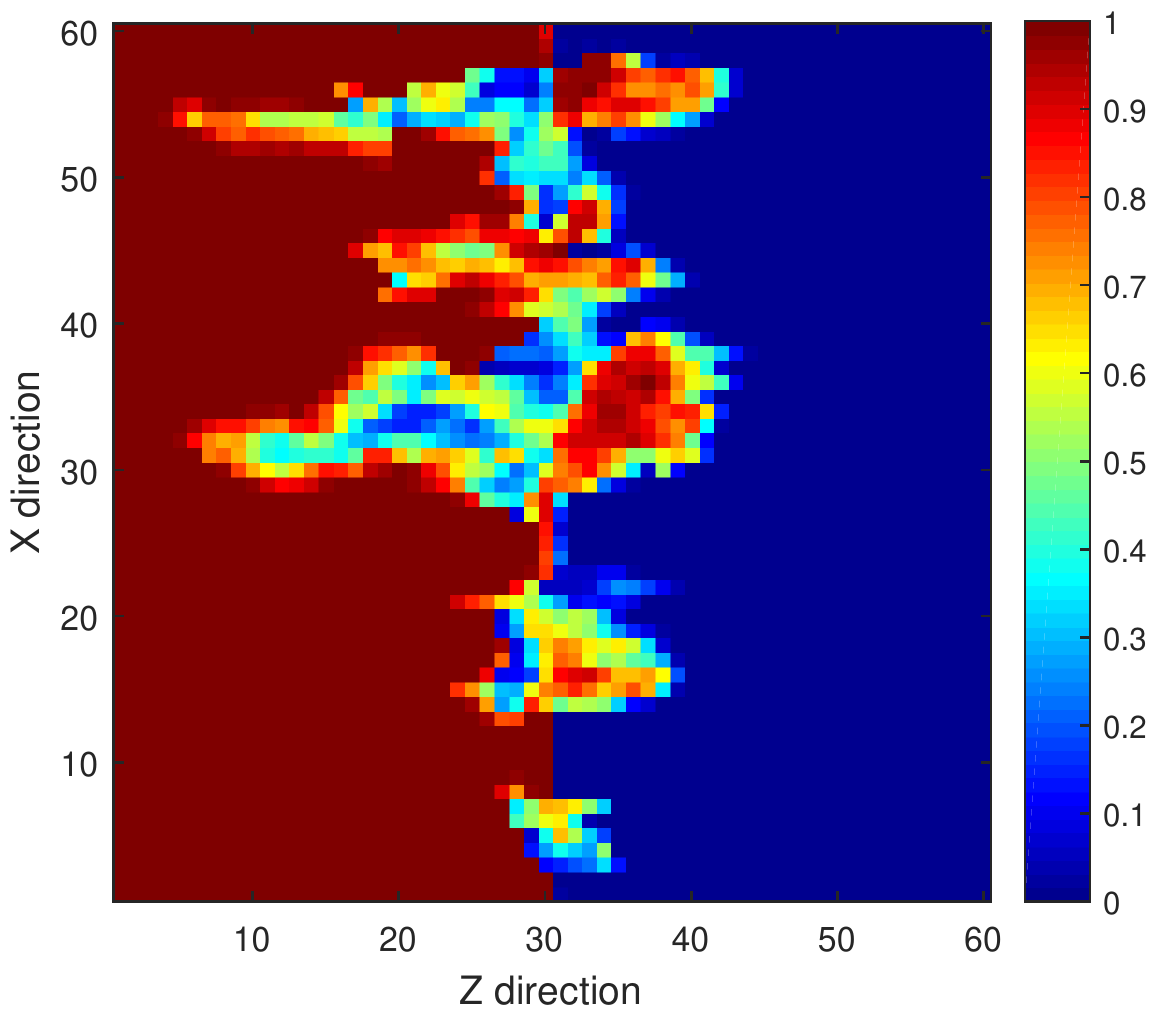}
\caption{Oil saturation profile of Case 1c }
\label{fig:s_he_se_pg}
\end{figure}

\begin{figure}[!htb]
\centering
\includegraphics[scale=0.6]{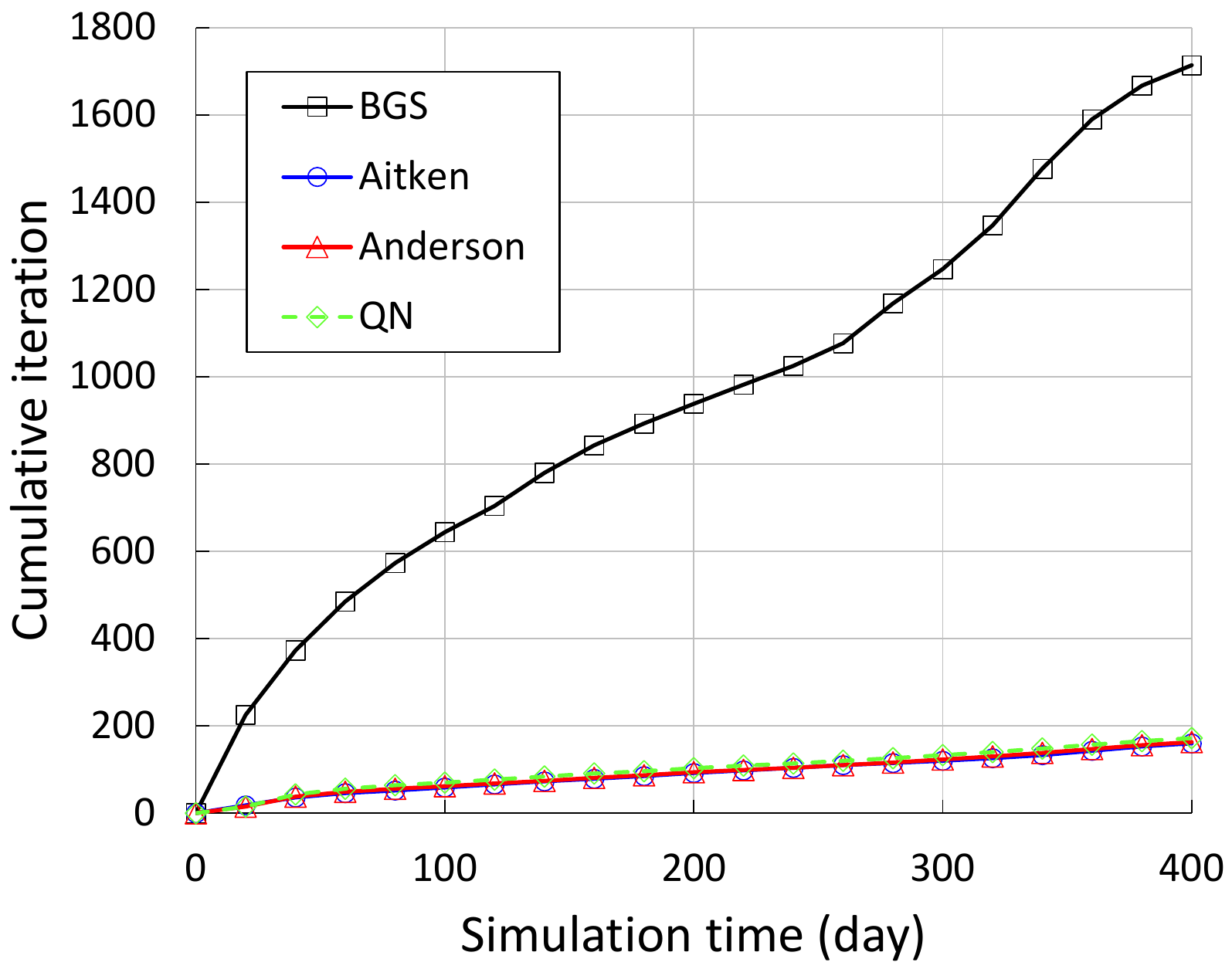}
\caption{Cumulative number of outer iterations versus simulation time of Case 1c }
\label{fig:cN_he_se_pg}
\end{figure}

We also test a counter-current flow problem on the heterogeneous square model. The gravity is towards the Z direction. In this case, complex fluid dynamics quickly develop due to gravity segregation after the beginning of the simulation. The oil saturation profile and the cumulative number of outer iterations versus simulation time for the IHU flux are plotted in \textbf{Fig. \ref{fig:s_he_se_pg}} and \textbf{Fig. \ref{fig:cN_he_se_pg}}, respectively. It can be seen that the presence of the counter-current flow further exacerbates the nonlinear convergence difficulty encountered by BGS. The three acceleration methods achieve reduction in the total iterations by more than an order of magnitude. We do not plot the cumulative iterations for the PPU flux, because BGS exhibits an even worse convergence performance with PPU.

\subsubsection{Summary of outer iteration performance }

The outer iteration performance of Case 1 is summarized in Table \ref{tab:iter_1}. We can see that the basic BGS process leads to huge numbers of outer iterations and wasted computations. In contrast, the three acceleration methods provide much faster convergence. The three methods are almost equivalent showing similar performance, though Anderson enjoys the least number of iterations overall. In the presence of buoyancy, highly heterogeneous permeability and porosity fields, and sharp saturation fronts propagating in the domain, the coupling between flow (\ref{eq:press_eq}) and transport (\ref{eq:Fl_tol_v}) becomes quite strong. The IHU scheme shows superior convergence performance compared with PPU. Especially in Case 1a, IHU achieves a significant iteration reduction for BGS. Nevertheless, it should be noted that BGS still exhibits severe convergence difficulties under the other cases which are heterogeneous and thus more challenging. Therefore we can conclude that IHU is capable of reducing the coupling degree between the sub-problems to some extent, yet could not resolve the related issues altogether. 

\begin{table}[!htb]
\centering
\caption{Outer iteration performance of Case 1 }
\label{tab:iter_1}
\begin{tabular}{|c|c|c|c|c|c|c|c|c|}
\hline
\multirow{2}{*}{CASE} & \multicolumn{2}{c|}{BGS} & \multicolumn{2}{c|}{Aitken} & \multicolumn{2}{c|}{Anderson} & \multicolumn{2}{c|}{QN} \\ \cline{2-9} 
                      & PPU         & IHU         & PPU          & IHU           & PPU           & IHU            & PPU        & IHU         \\ \hline
1a          & 234         & 47         & 53           & 42           & 49            & 39            & 49         & 39         \\ \hline
1b          & 1469        & 834        & 211          & 133          & 209           & 126           & 223        & 115        \\ \hline
1c          & 1969        & 1714       & 269          & 160          & 237           & 163           & 285        & 172        \\ \hline
\end{tabular}
\end{table}

In order to better understand the behaviors of the different numerical flux schemes and acceleration methods, we carry out detailed analysis on the coupling effects and mechanisms. For incompressible multiphase flow and transport in one dimension with a fixed total velocity, the saturation can be obtained by solving the transport problem only. In multi-dimensional domains, the assumption that the total velocity does not vary in time and space is no longer valid, and we have to solve the coupled flow and transport problem for each time-step. On one hand, the outer coupling in SFI results from the strong dependence of the saturation field on the total velocity computed from the pressure distribution. On the other hand, changes in the saturation alter the phase mobility terms, which in turn affect the pressure distribution. 

Buoyancy adds to the complexity by introducing flow reversal phenomenon during iterations. In the PPU scheme, the upwinding of a fluid phase is determined according to the phase potential difference at cell interface. Recent studies have revealed that the non-differentiability in PPU due to the upstream direction switching could be a major cause of nonlinear convergence difficulties. We demonstrate PPU for evaluating the mobility under a simplified two-phase (oil and water) scenario. The total velocity is taken to be positive and fixed. We first define the dimensionless gravity number $C_g$ as
\begin{equation}
C_g = \frac{kg(\rho_w-\rho_o)}{\mu_o}\nabla h
\end{equation}
From Eq. (\ref{eq:Fl_tol_v}), we can rewrite the upstream conditions (\ref{eq:PPU}) in terms of $u_T$ as
\begin{equation*} 
\lambda_{w,ij}^{PPU} = \left\{ {\begin{array}{*{20}c}
\lambda_{w}(s_{i}),   &   u_T + C_g k_{ro} > 0 , \\ 
\lambda_{w}(s_{j}), &  \mathrm{otherwise} .
\end{array}} \right.
\end{equation*}
and
\begin{equation} 
\label{Eq:up_c}
\lambda_{o,ij}^{PPU} = \left\{ {\begin{array}{*{20}c}
\lambda_{o}(s_{i}), &   u_T - C_g M k_{rw} > 0 , \\ 
\lambda_{o}(s_{j}), &  \mathrm{otherwise} .
\end{array}} \right.
\end{equation}
Brenier and Jaffre (1991) showed how to explicitly determine the upstream direction of $\lambda_{l}$ using Eq. (\ref{Eq:up_c})
\begin{equation} 
\lambda_{w,ij}^{PPU} = \left\{ {\begin{array}{*{20}c}
\lambda_{w}(s_{i}), &  \theta_w > 0 , \\ 
\lambda_{w}(s_{j}), &  \mathrm{otherwise} .
\end{array}} \right.
\end{equation}
where
\begin{equation} 
\label{Eq:theta_w}
\theta_w =  u_T + C_g k_{ro}(s_i) 
\end{equation}
and
\begin{equation} 
\lambda_{o,ij}^{PPU} = \left\{ {\begin{array}{*{20}c}
\lambda_{o}(s_{i}), &  \theta_o > 0 , \\ 
\lambda_{o}(s_{j}), &  \mathrm{otherwise} .
\end{array}} \right.
\end{equation}
where
\begin{equation} 
\label{Eq:theta_o}
\theta_o = u_T - C_g M k_{rw}(s_i)
\end{equation}
As can be seen, flow reversal from co-current to counter-current regime at a cell interface depends only on the saturation change in the upwind cell. From Eqs. (\ref{Eq:theta_w}) and (\ref{Eq:theta_o}), if $C_g < 0$, $\theta_o$ is always positive, and subsequently
\begin{equation} 
\label{Eq:kr_up}
\renewcommand*{\arraystretch}{1.3}
\left\{ {\begin{array}{*{20}c}
\lambda_{w,ij}^{PPU} = \lambda_{w}(s_{i}) \ \ \mathrm{and} \ \ \lambda_{o,ij}^{PPU} = \lambda_{o}(s_{i}), & 0\leq \theta_w \leq \theta_o \\ 
\lambda_{w,ij}^{PPU} = \lambda_{w}(s_{j}) \ \ \mathrm{and} \ \ \lambda_{o,ij}^{PPU} = \lambda_{o}(s_{i}), & \theta_w \leq 0 \leq \theta_o 
\end{array}} \right.
\end{equation}
If $C_g > 0$, now $\theta_w$ is always positive, and thus
\begin{equation} 
\label{Eq:kr_up_2}
\renewcommand*{\arraystretch}{1.3}
\left\{ {\begin{array}{*{20}c}
\lambda_{w,ij}^{PPU} = \lambda_{w}(s_{i}) \ \ \mathrm{and} \ \ \lambda_{o,ij}^{PPU} = \lambda_{o}(s_{i}), & 0 \leq \theta_o \leq \theta_w \\ 
\lambda_{w,ij}^{PPU} = \lambda_{w}(s_{i}) \ \ \mathrm{and} \ \ \lambda_{o,ij}^{PPU} = \lambda_{o}(s_{j}), & \theta_o \leq 0 \leq \theta_w
\end{array}} \right.
\end{equation}

The flow reversal in the PPU flux can lead to cycling or divergence of the Newton iterative process, and thus largely reduce the solution efficiency of the transport sub-problem. Moreover, this discontinuous behavior also has significant impacts on the convergence of the outer loop within the SFI framework. According to Eqs. (\ref{Eq:theta_w}) and (\ref{Eq:theta_o}), variation of the total flux may induce the upstream direction switching in the transport equations, if the coupling effects are strong. Specifically, the upstream direction may change after solving for the saturation; then it may change again when the saturation is updated based on the pressure and total flux for the next outer iteration.

The IHU flux is recently proposed to address the flow reversal issue. The newton solver for the transport sub-problem obviously will benefit from the continuously differentiable property of IHU. Moreover, the flow reversal within the transport equations due to the variation of the total flux with the same sign can be avoided. However, IHU is non-differentiable along the surface defined by $u_{T,ij} \left ( \Delta p_{ij}, s_i, s_j \right ) = 0$, according to Eq. (\ref{eq:V_HU}). Indeed, the upwinding direction will switch inevitably, if the sign of the total flux obtained from the pressure equation changes.

Hamon et al. (2016) develop a differentiable total-velocity scheme which improves the nonlinear convergence of fully-coupled flow and transport problem. But the sign of total flux can still change back and forth between the sequential solutions from one outer iteration to the next. It is expected that the `flip-flopping' issue due to total flux is a primary factor affecting the outer-loop convergence. Therefore, PPU is simply applied for the total velocity discretization in this work.

\begin{figure}[!htb]
\centering
\includegraphics[scale=0.7]{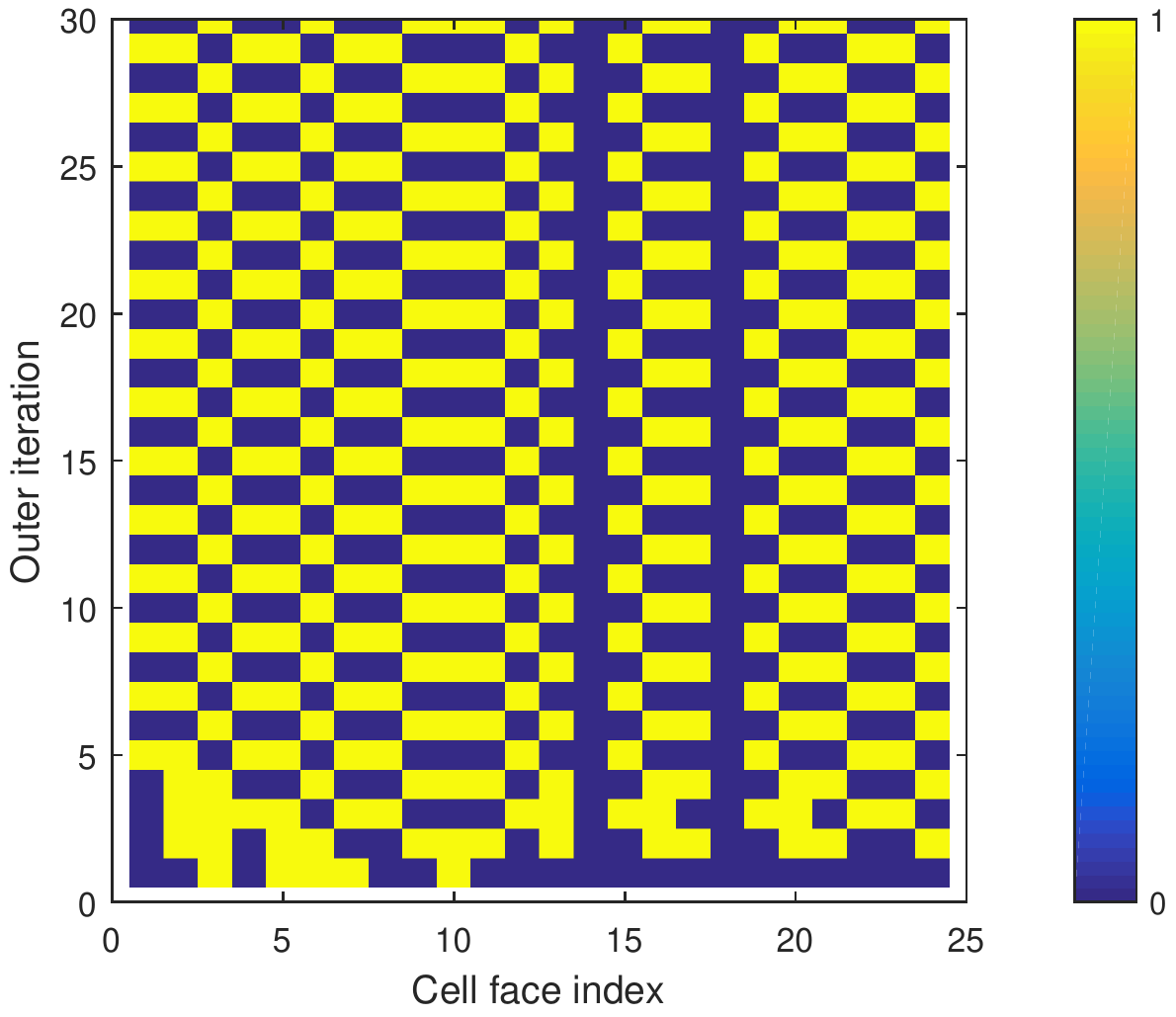}
\caption{Sign changes of the total flux across each interface during the basic BGS process }
\label{fig:4_by_4_square}
\end{figure}

To illustrate the above aspect, we use a $4 \times 4$ sub-domain taken from the heterogeneous square model in Case 1c. The initial condition becomes $s = 0.5$. The time-step size is set to 1000 days. In \textbf{Fig. \ref{fig:4_by_4_square}}, we plot the sign changes of the total flux across each interface during the basic BGS process. It can be seen that the flux directions at most of the interfaces keep switching between iterations, causing severe convergence problem of the outer loop. This explains why the performance of IHU degrades substantially in Cases 1b and 1c. The three nonlinear acceleration techniques, on the other hand, are surprisingly effective to stabilize the iterations, even when PPU is applied.

\subsection{Case 2: two-phase, pure gravitational, SPE 10 model }

\begin{figure}[!htb]
\centering
\subfloat[Permeability (md)]{
\includegraphics[scale=0.61]{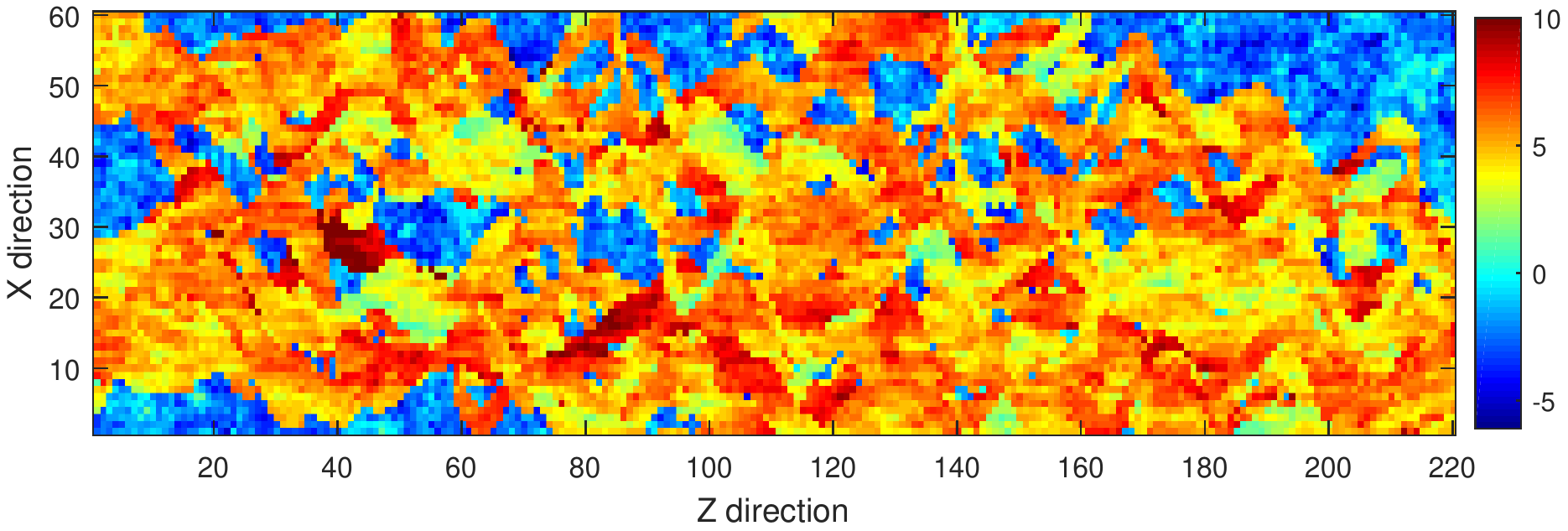}}
\\
\subfloat[Porosity]{
\includegraphics[scale=0.61]{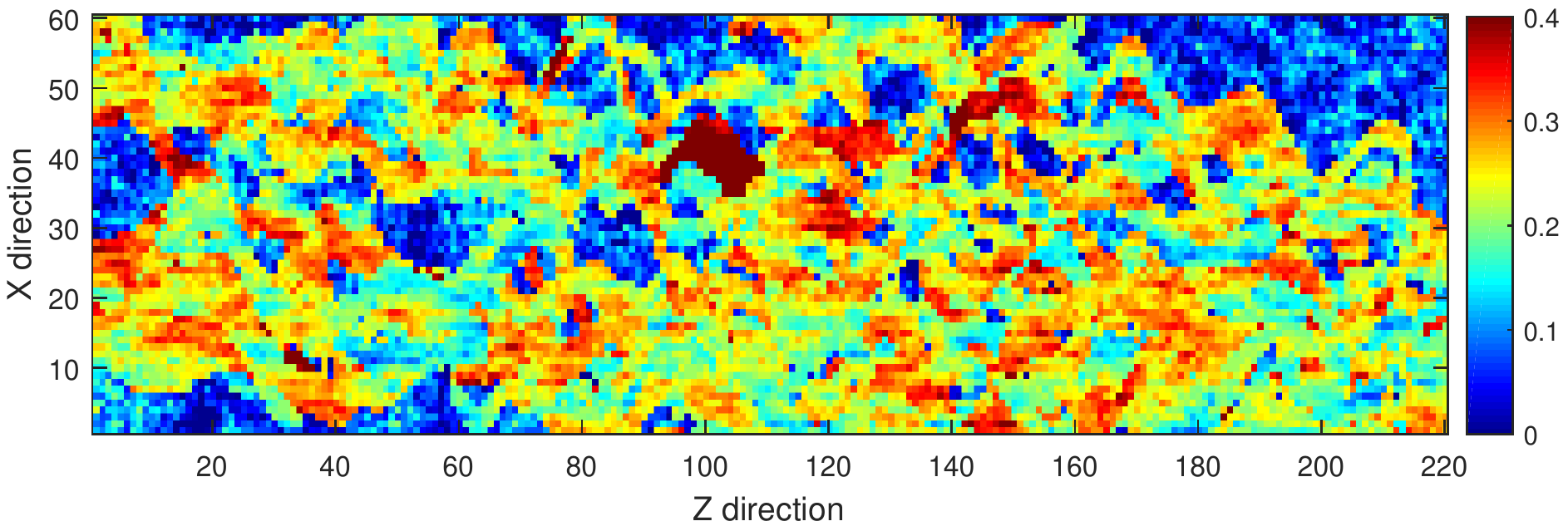}}
\caption{Permeability (log) and porosity fields of the bottom layer in the SPE 10 model}
\label{fig:perm_poro}
\end{figure}

For this numerical case, the model setting is identical to the previous case as specified in Table \ref{tab:specification}, except that the bottom layer of the SPE 10 model is used. The rock properties are shown in \textbf{Fig. \ref{fig:perm_poro}}. Also time-interval size becomes 20 days, with total simulation time as 400 days.

\subsubsection{Case 2a }

\begin{figure}[!htb]
\centering
\includegraphics[scale=0.6]{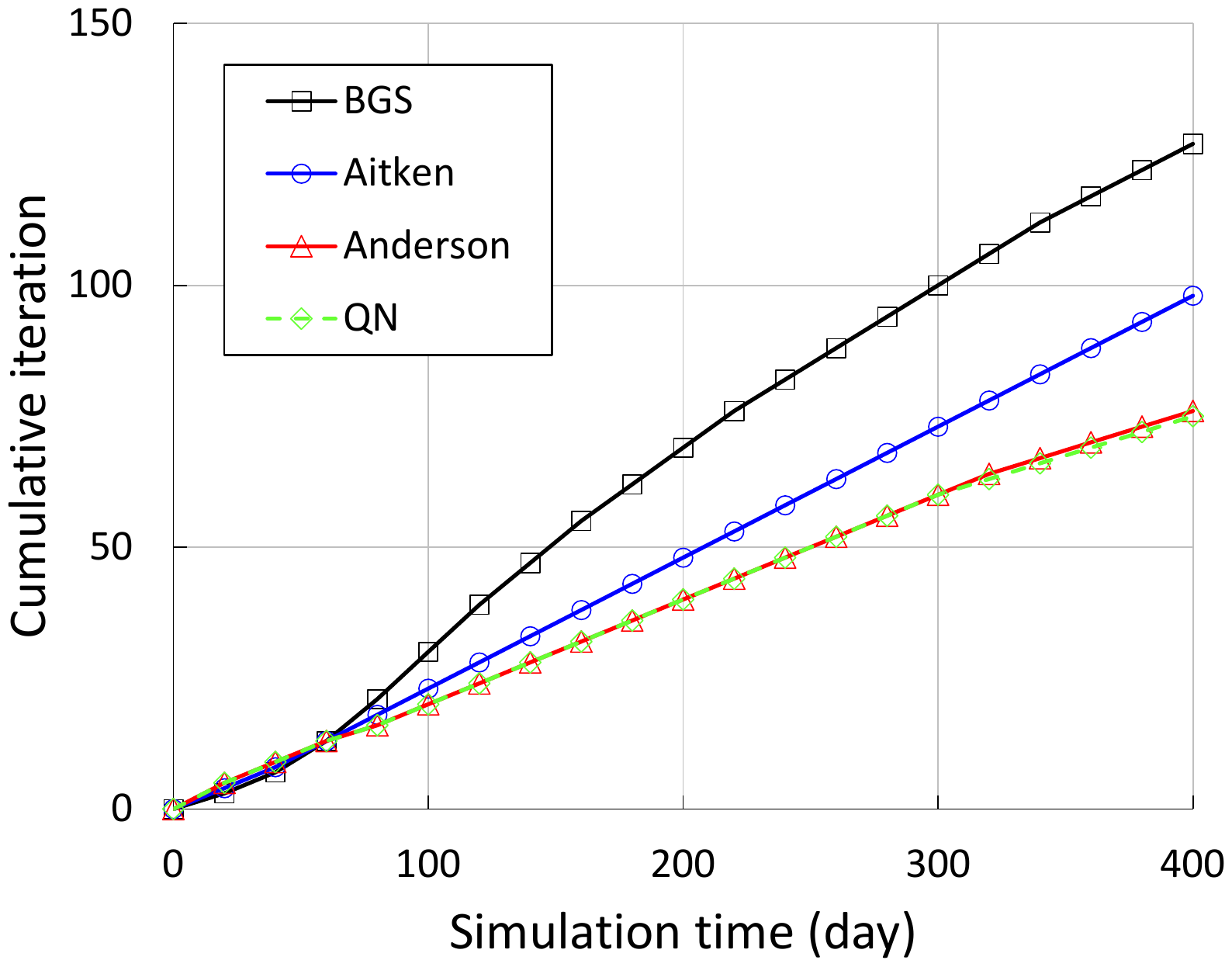}
\caption{Cumulative number of outer iterations versus simulation time of Case 2a }
\label{fig:cN_2_ne_spe_pg}
\end{figure}

Initial water saturation is uniformly set to 0.1. The cumulative number of outer iterations versus simulation time for the IHU flux is plotted in \textbf{Fig. \ref{fig:cN_2_ne_spe_pg}}. Due to the non-equilibrium initial condition, gravity segregation starts to take place in all cells at time zero. We can see that the nonlinear acceleration methods reduce the outer iterations required by the BGS process. For this particular case, Aitken performs worse than QN and Anderson, which incorporate more information of the previous iterations to guide the current update.

\subsubsection{Case 2b }

\begin{figure}[!htb]
\centering
\includegraphics[scale=0.61]{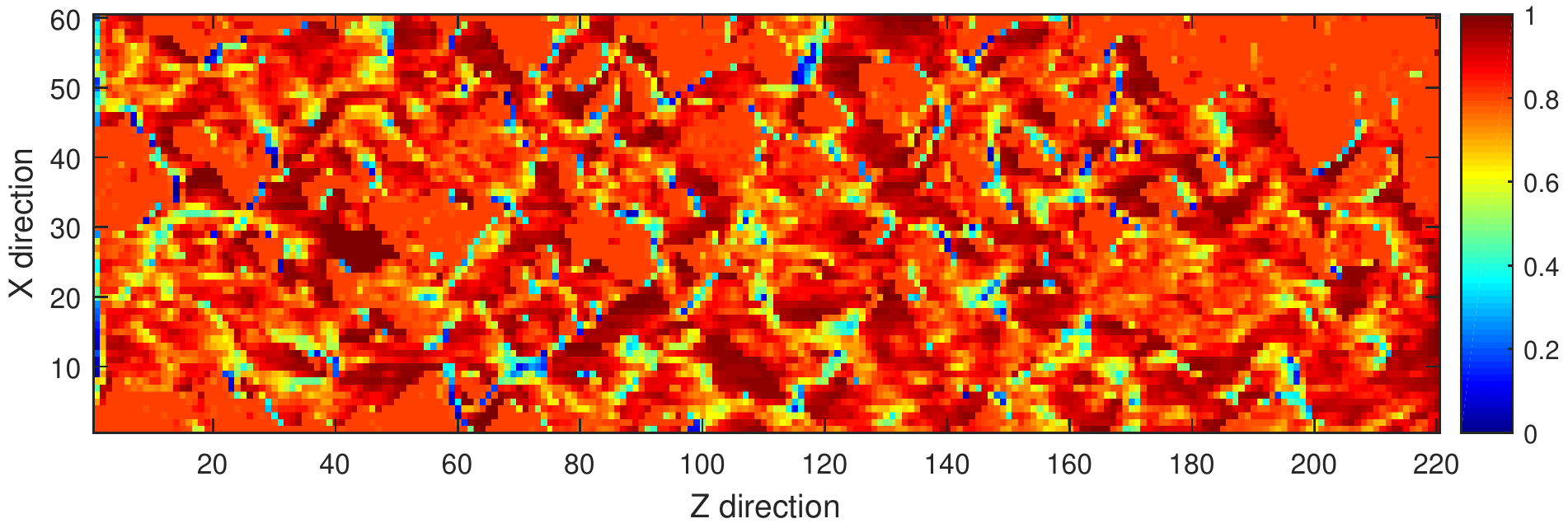}
\caption{Oil saturation profile of Case 2b }
\label{fig:s_ne_spe_pg}
\end{figure}

\begin{figure}[!htb]
\centering
\includegraphics[scale=0.6]{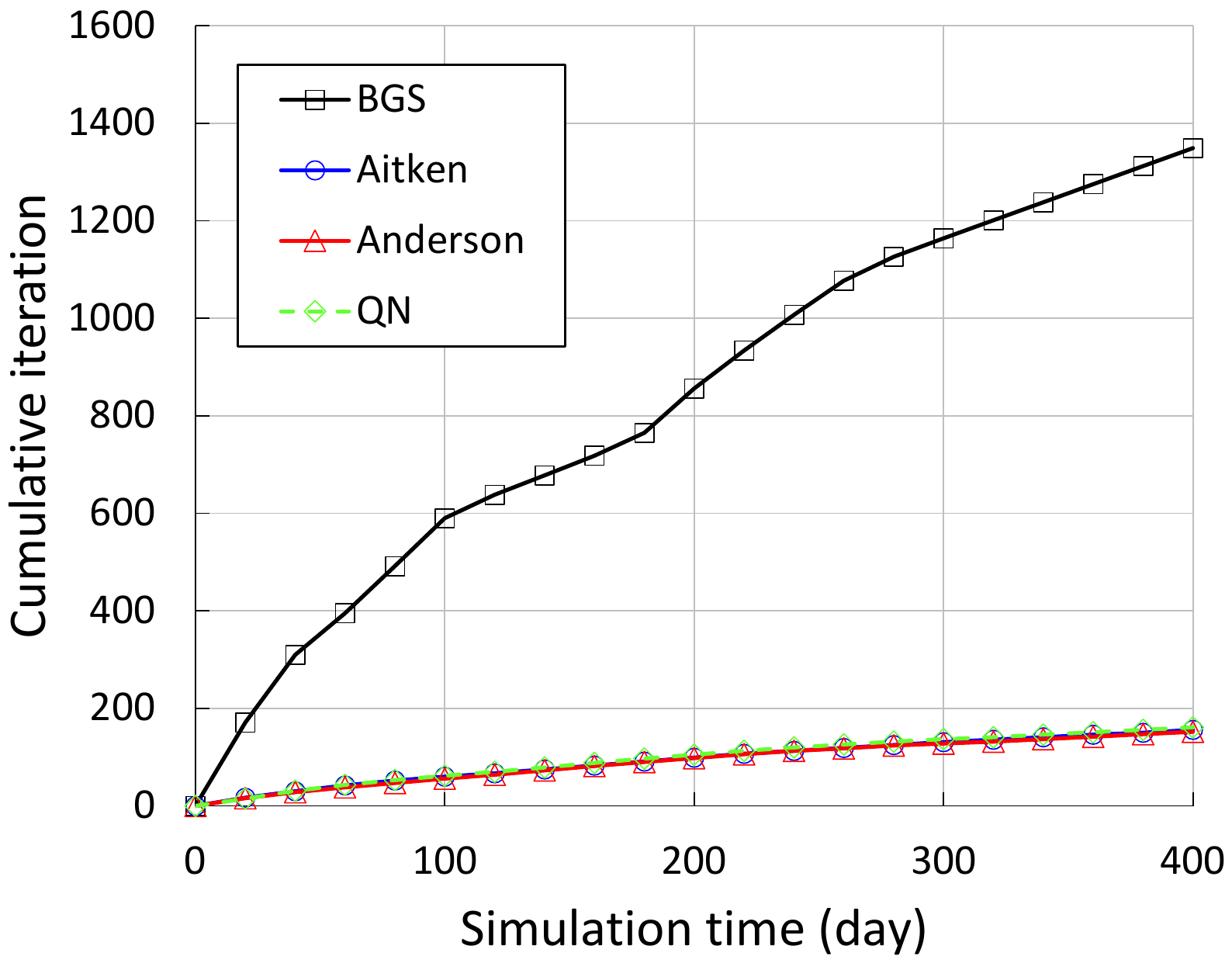}
\caption{Cumulative number of outer iterations versus simulation time of Case 2b }
\label{fig:cN_1_ne_spe_pg}
\end{figure}

Initial water saturation is changed to 0.2. The oil saturation profile and the cumulative number of outer iterations versus simulation time for the IHU flux are plotted in \textbf{Fig. \ref{fig:s_ne_spe_pg}} and \textbf{Fig. \ref{fig:cN_1_ne_spe_pg}}, respectively. Compared to Case 2a, the higher initial saturation results in a higher fluid mobility. This translates into strong coupling effects between flow and transport. Therefore BGS undergoes a remarkable deterioration in the performance, leading to a large number of time-step cuts.

\subsubsection{Summary of outer iteration performance }

The outer iteration performance of Case 2 is summarized in Table \ref{tab:iter_2}. The results show the superior convergence properties of the IHU scheme and the acceleration methods.

\begin{table}[!htb]
\centering
\caption{Outer iteration performance of Case 2 }
\label{tab:iter_2}
\begin{tabular}{|c|c|c|c|c|c|c|c|c|}
\hline
\multirow{2}{*}{CASE} & \multicolumn{2}{c|}{BGS} & \multicolumn{2}{c|}{Aitken} & \multicolumn{2}{c|}{Anderson} & \multicolumn{2}{c|}{QN} \\ \cline{2-9} 
                      & PPU         & IHU         & PPU          & IHU           & PPU           & IHU            & PPU        & IHU         \\ \hline
2a      & 687         & 127        & 102          & 98           & 92            & 76            & 92         & 75         \\ \hline
2b      & 1608        & 1349       & 193          & 156          & 179           & 152           & 186        & 161        \\ \hline
\end{tabular}
\end{table}

\subsection{Case 3: two-phase, viscous and gravitational forces, SPE 10 model }

We consider the scenario with combined viscous and gravitational forces. A quarter-five spot well pattern is applied: water is injected at the middle of the domain and producers are placed at the four corners. The injection rate is 2.5e-4 Pore Volume (18.81 $\textrm{m}^3$) per day. The modified model parameters of Case 3 are summarized in Table \ref{tab:specification_VG}. The other parameters specified in Case 1a remain unchanged.

\begin{table}[!htb]
\centering
\caption{Modified model parameters of Case 3 }
\label{tab:specification_VG}
\begin{tabular}{|c|c|c|}
\hline
Parameter                  &  Value            & Unit      \\ \hline
NX / NZ                    &  60 / 220         &           \\ \hline
LX / LY / LZ               &  600 / 10 / 2200  & ft        \\ \hline
Initial water saturation   &  0.01             &           \\ \hline
Injection rate             &  18.81           & $\textrm{m}^3/\textrm{d}$  \\ \hline
Production BHP             &  500             & psi          \\ \hline
Time-interval size         &  20            & day    \\ \hline
Total simulation time      &  200           & day    \\ \hline


\end{tabular}
\end{table}

\subsubsection{Case 3a }

\begin{figure}[!htb]
\centering
\includegraphics[scale=0.61]{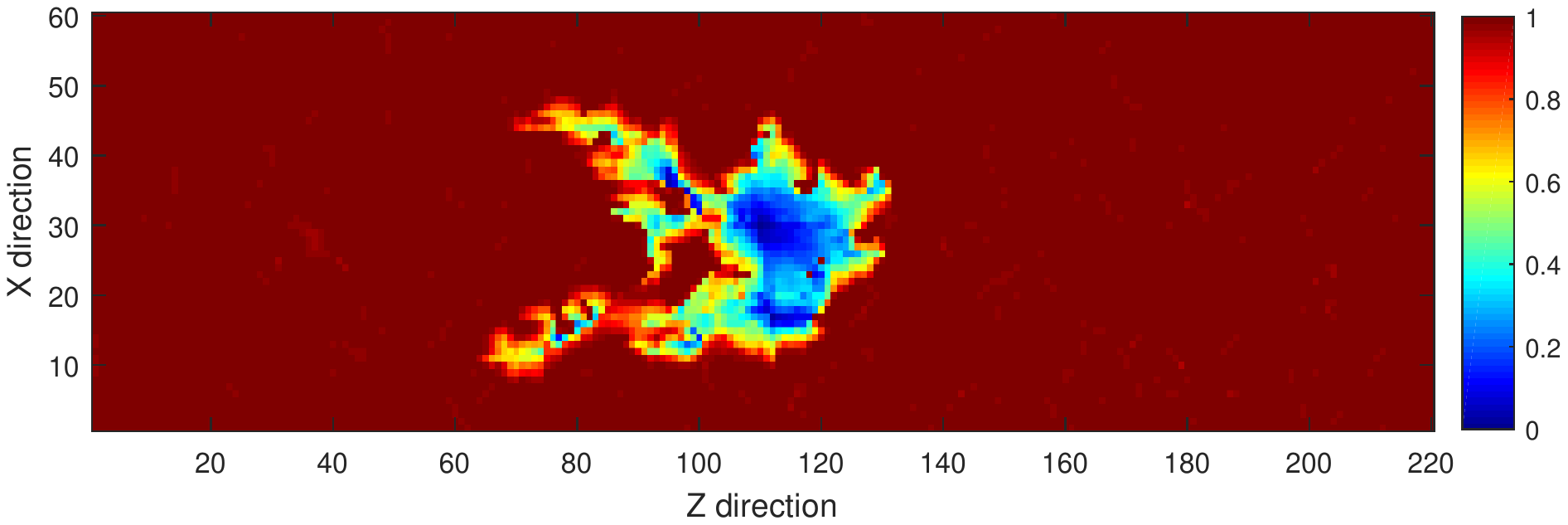}
\caption{Oil saturation profile of Case 3a }
\label{fig:s_VG_1_spe}
\end{figure}

\begin{figure}[!htb]
\centering
\includegraphics[scale=0.6]{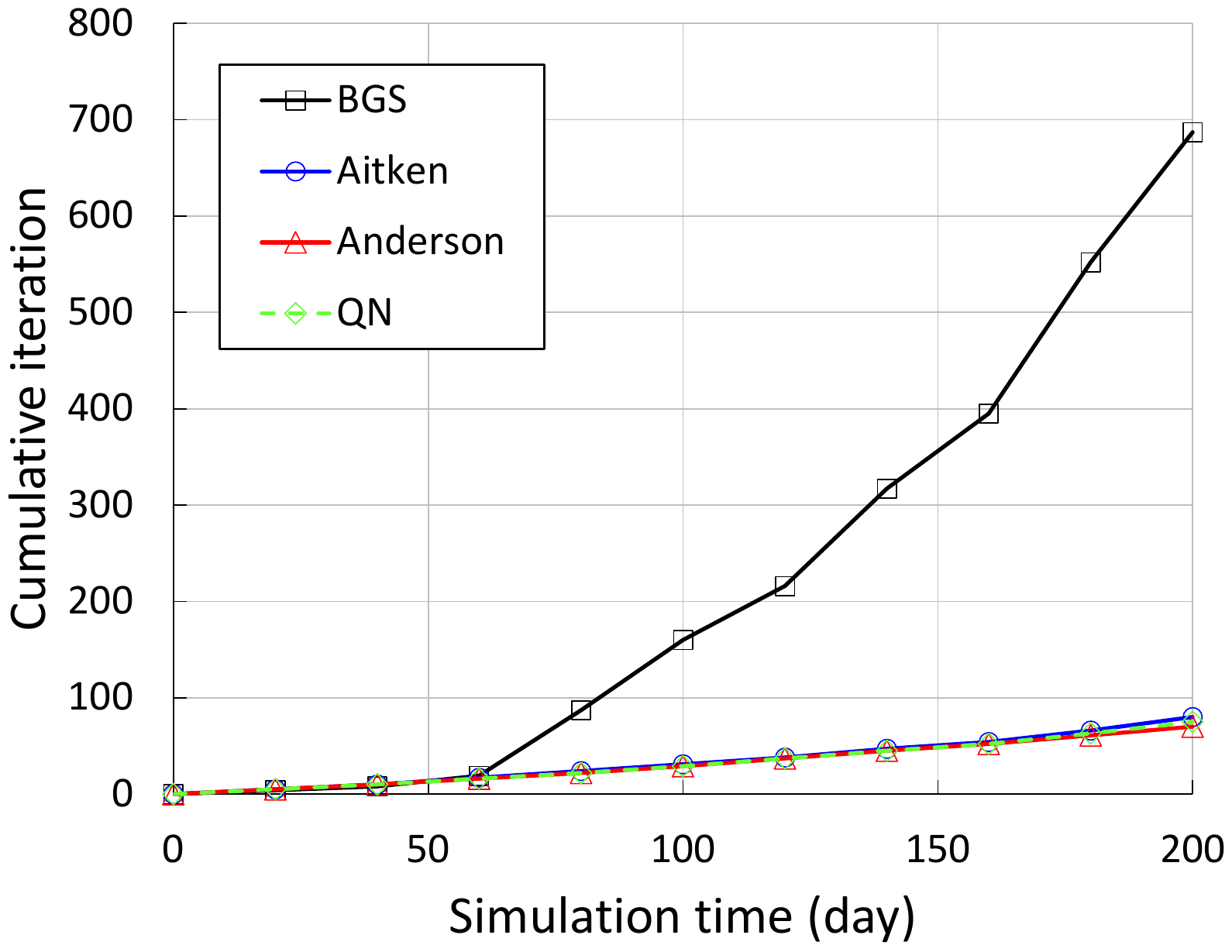}
\caption{Cumulative number of outer iterations versus simulation time of Case 3a }
\label{fig:cN_VG_1_spe}
\end{figure}

The oil saturation profile and the cumulative number of outer iterations versus simulation time for the IHU flux are plotted in \textbf{Fig. \ref{fig:s_VG_1_spe}} and \textbf{Fig. \ref{fig:cN_VG_1_spe}}, respectively. It can be seen that BGS encounters severe convergence difficulties after an early period. In contrast, the nonlinear acceleration methods show a smooth convergence behavior.

The outer iteration performance of Case 3a is summarized in Table \ref{tab:iter_3}. With the basic BGS process, the flow scenario under combined viscous and gravitational forces is quite challenging for both the PPU and IHU schemes. Anderson together with IHU results in the least number of iterations.

\begin{table}[!htb]
\centering
\caption{Outer iteration performance of Case 3a }
\label{tab:iter_3}
\begin{tabular}{|l|l|l|l|l|l|l|l|l|}
\hline
\multirow{2}{*}{} & \multicolumn{2}{l|}{BGS} & \multicolumn{2}{l|}{Aitken} & \multicolumn{2}{l|}{Anderson} & \multicolumn{2}{l|}{QN} \\ \cline{2-9} 
                  & PPU         & IHU         & PPU           & IHU          & PPU            & IHU           & PPU         & IHU        \\ \hline
SUM               & 908         & 687        & 109           & 80          & 113            & 70           & 139         & 75        \\ \hline
\end{tabular}
\end{table}

\subsubsection{Case 3b }

\begin{figure}[!htb]
\centering
\includegraphics[scale=0.61]{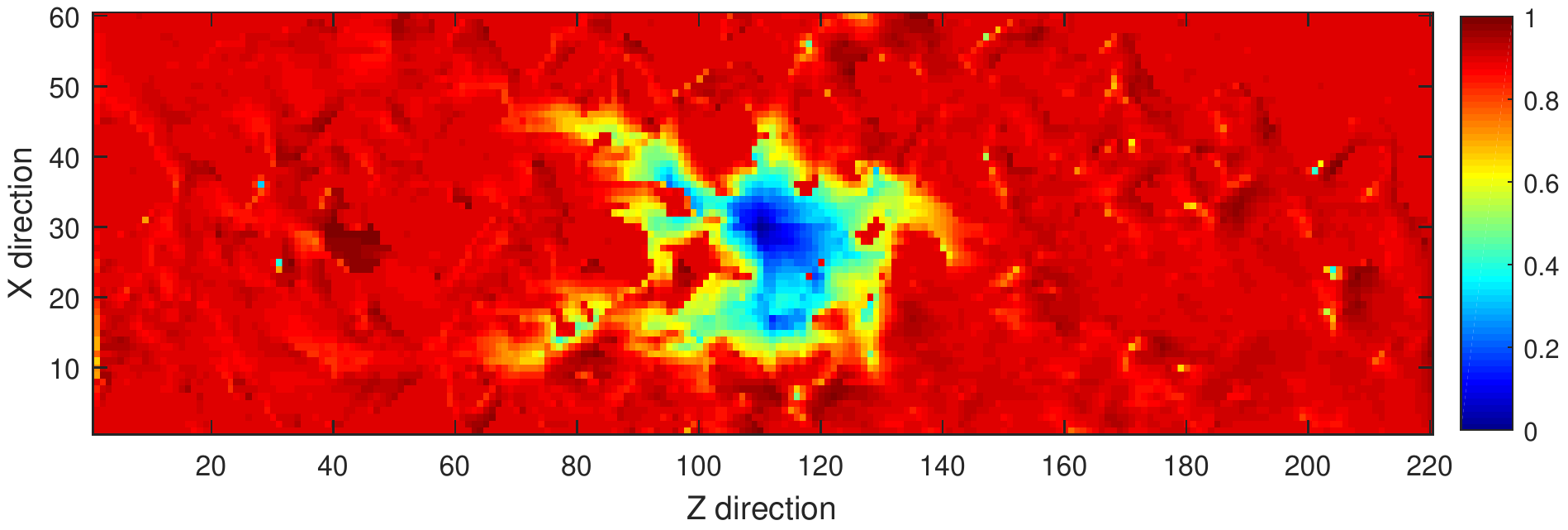}
\caption{Oil saturation profile of Case 3b }
\label{fig:s_VG_2_spe}
\end{figure}

\begin{figure}[!htb]
\centering
\includegraphics[scale=0.6]{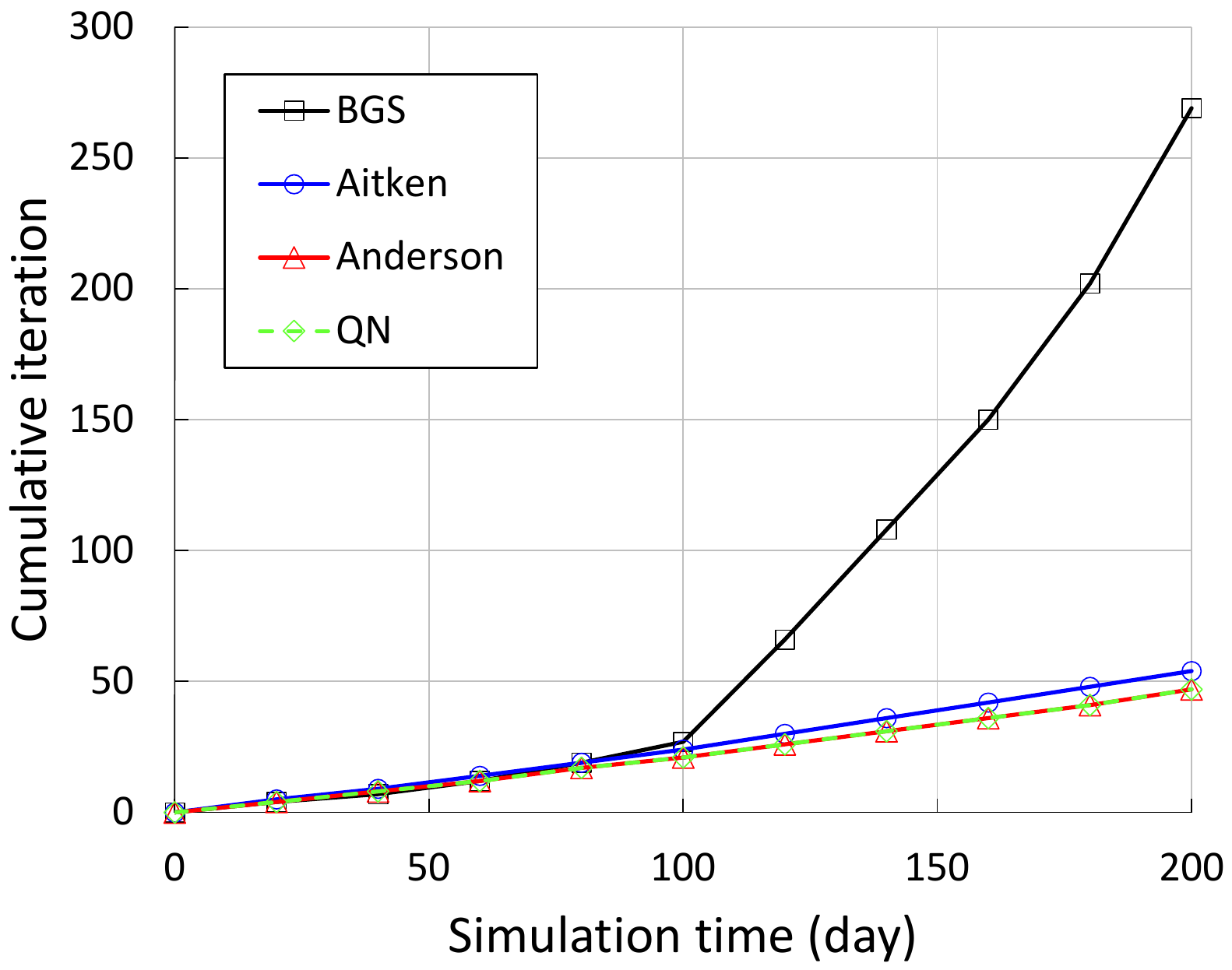}
\caption{Cumulative number of outer iterations versus simulation time of Case 3b }
\label{fig:cN_VG_2_spe}
\end{figure}

For this case, oil density is changed to 800 $\textrm{kg}/\textrm{m}^3$, and initial water saturation becomes 0.1. The oil saturation profile and the cumulative number of outer iterations versus simulation time for the IHU flux are plotted in \textbf{Fig. \ref{fig:s_VG_2_spe}} and \textbf{Fig. \ref{fig:cN_VG_2_spe}}, respectively.

\begin{figure}[!htb]
\centering
\includegraphics[scale=0.7]{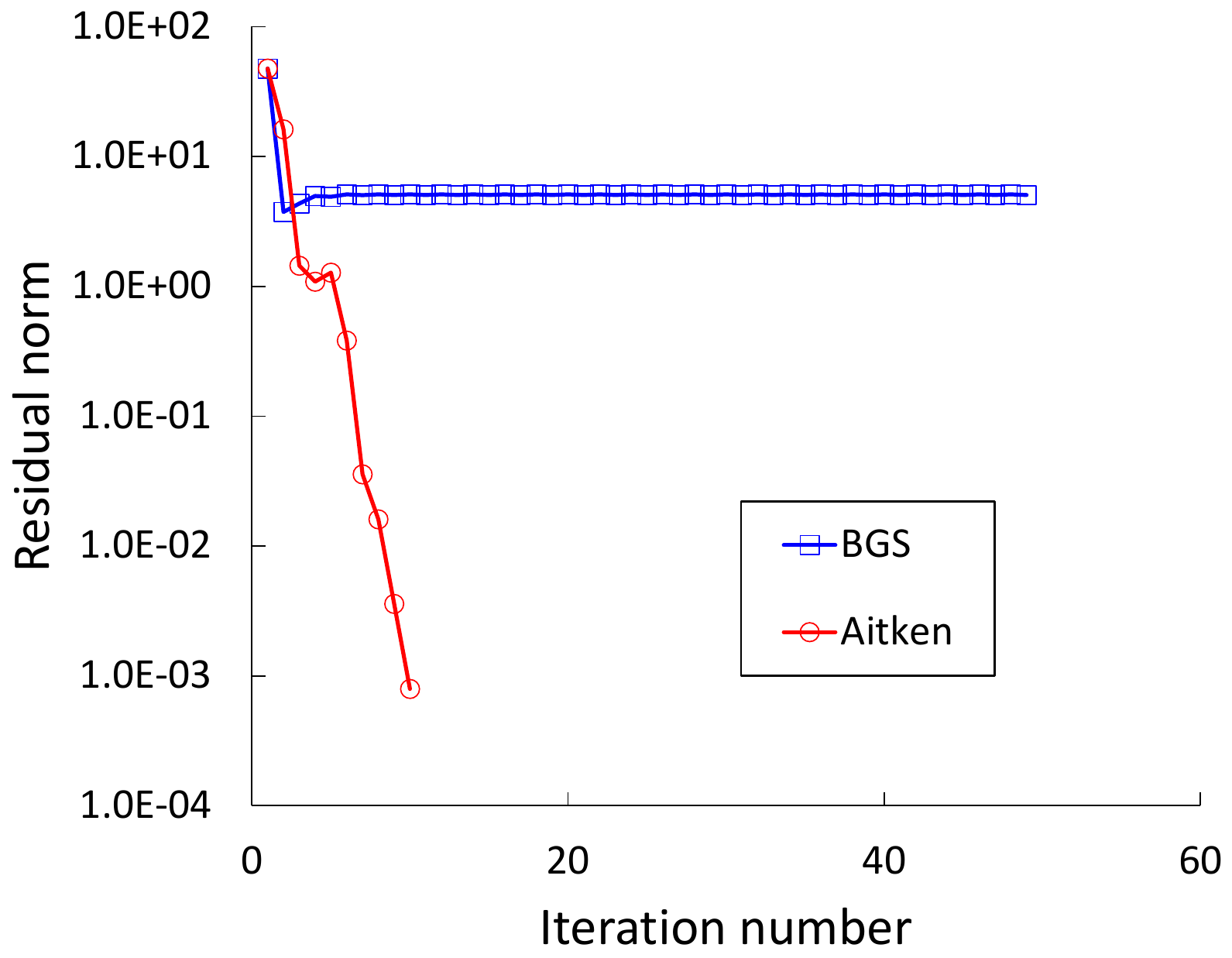}
\caption{Residual history of the time-step size as 100 days for Case 3b }
\label{fig:rh_VG_2_spe}
\end{figure}

We plot the residual history of the time-step size as 100 days in \textbf{Fig. \ref{fig:rh_VG_2_spe}}. We can clearly see that the BGS iterations completely stagnate. A closer look at the solution states during the iterations reveals that the updates oscillate between the flow and transport sub-problems. This is because the sign of the total flux changes as a function of iteration, causing the corresponding switching of the upwinding direction. The flow reversal phenomenon has a detrimental effect that prevents the BGS process from converging. In comparison to BGS, the numerical relaxation by Aitken provides a monotonous and steep residual decrease. Aitken enjoys an interesting mechanism to circumvent the oscillations on the outer level.

\subsection{Case 4: two-phase, pure viscous, SPE 10 model }

\subsubsection{Case 4a }

\begin{figure}[!htb]
\centering
\includegraphics[scale=0.61]{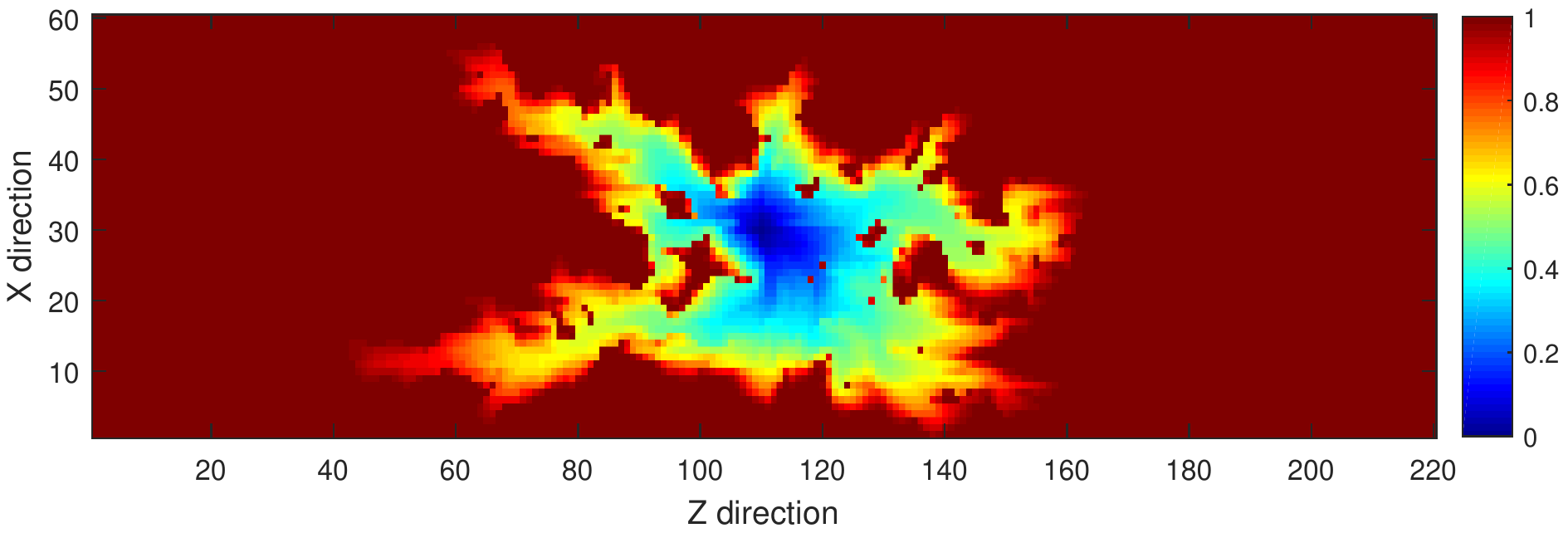}
\caption{Oil saturation profile of Case 4a }
\label{fig:s_V_1_spe}
\end{figure}

\begin{figure}[!htb]
\centering
\includegraphics[scale=0.6]{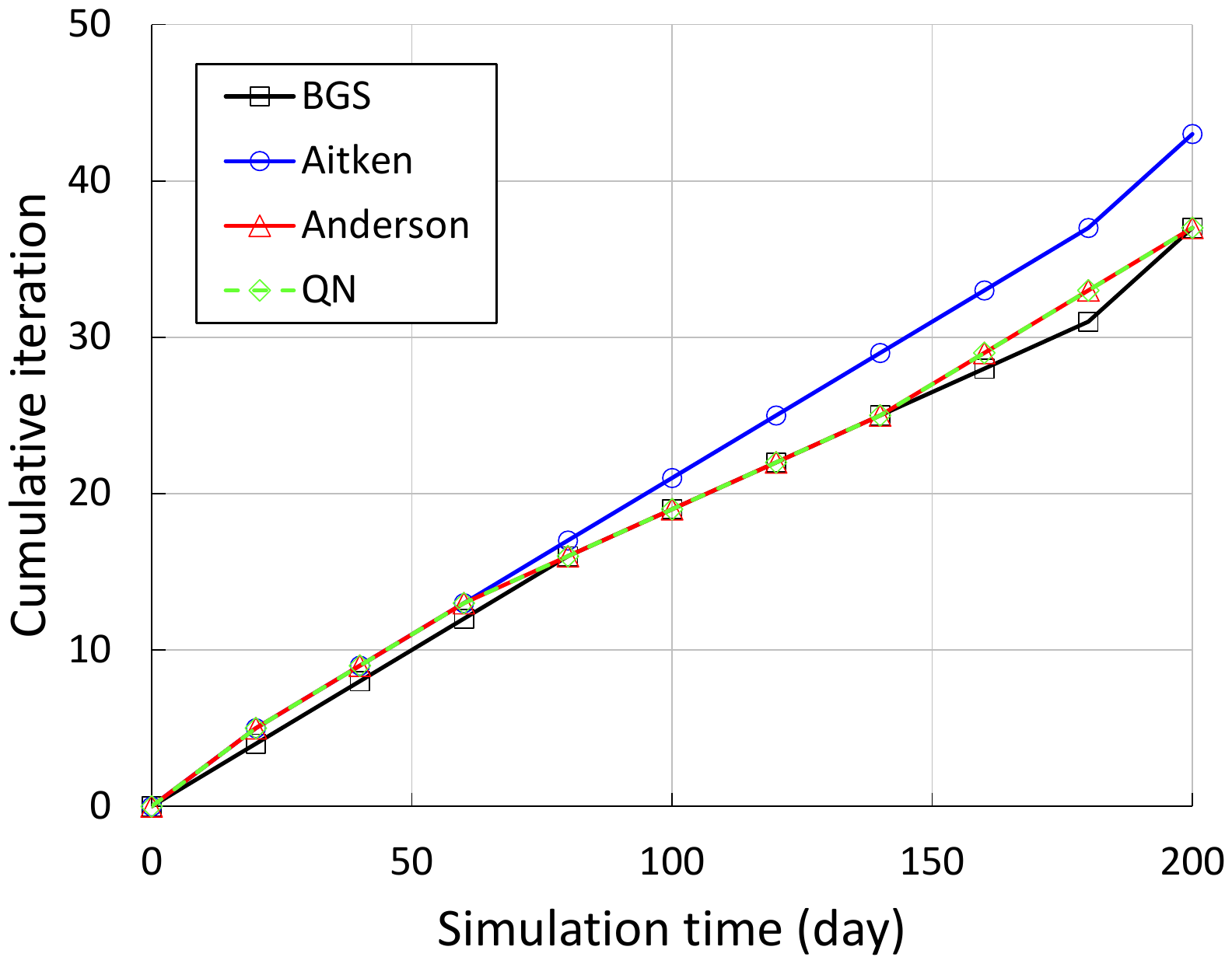}
\caption{Cumulative number of outer iterations versus simulation time of Case 4a }
\label{fig:cN_V_1_spe}
\end{figure}

We consider the scenario with pure viscous force. The model parameters of Case 3a are used. The injection rate is changed to 37.61 $\textrm{m}^3/\textrm{d}$. The oil saturation profile and the cumulative number of outer iterations versus simulation time are plotted in \textbf{Fig. \ref{fig:s_V_1_spe}} and \textbf{Fig. \ref{fig:cN_V_1_spe}}, respectively.

In this simulation case, the basic BGS process shows a satisfying convergence performance, indicating weak coupling effects between the sub-problems. On the other hand, Aitken induces a slightly higher iteration count. This is due to the under-relaxation factor applied for the first outer iteration. Overall, the coupled problem is relatively easy to solve, and all the methods present comparable performance.

\subsubsection{Case 4b }

\begin{figure}[!htb]
\centering
\includegraphics[scale=0.61]{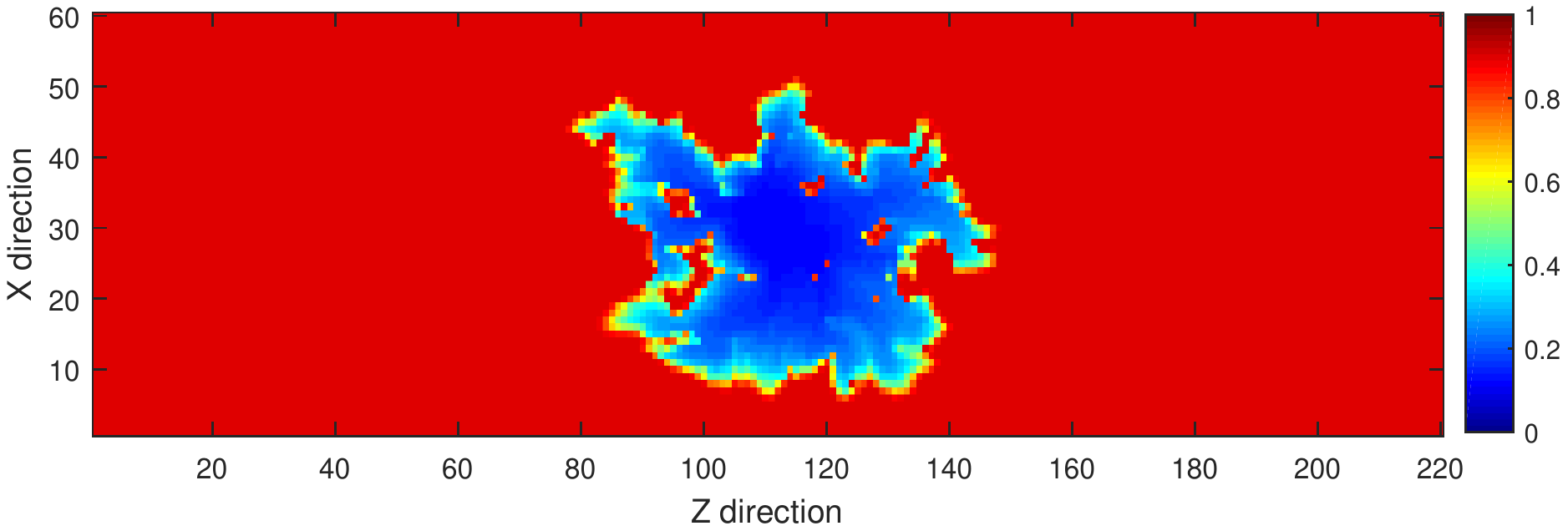}
\caption{Gas saturation profile of Case 4b }
\label{fig:s_WiG_spe}
\end{figure}

\begin{figure}[!htb]
\centering
\includegraphics[scale=0.6]{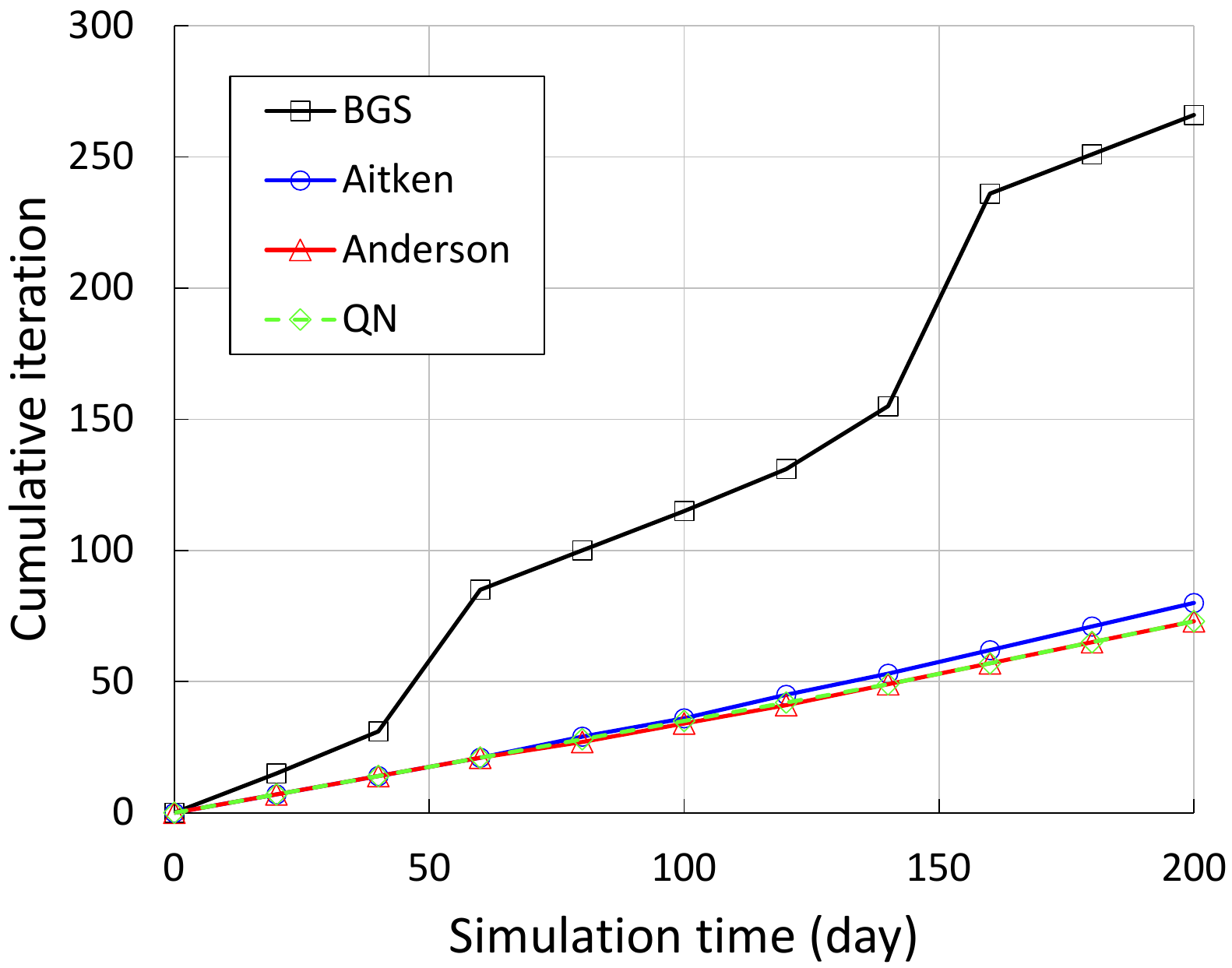}
\caption{Cumulative number of outer iterations versus simulation time of Case 4b }
\label{fig:cN_WiG_spe}
\end{figure}

We now test a model with water injecting into a gas reservoir. The specification of the model is shown in Table \ref{tab:specification_WiG}. Cubic relative-permeability functions are used. The oil saturation profile and the cumulative number of outer iterations versus simulation time are plotted in \textbf{Fig. \ref{fig:s_WiG_spe}} and \textbf{Fig. \ref{fig:cN_WiG_spe}}, respectively.

Comparing to Case 4a, the convergence performance of BGS degrades significantly. The poor convergence should not be resulted from the flow reversal issues, because gravitational force is neglected in the model. From the figure, we clearly observe that the saturation front is very sharp, due to the property of the fractional flow function. The propagation of the sharp front produces a tight coupling between the sub-problems: the pressure equation becomes a strong function of the mobility terms. The three acceleration methods provide a uniform convergence without any time-step cut.

\begin{table}[!htb]
\centering
\caption{Specification of the model with water injecting into a gas reservoir}
\label{tab:specification_WiG}
\begin{tabular}{|c|c|c|}
\hline
Parameter                  &  Value            & Unit      \\ \hline
NX / NZ                    &  60 / 220         &           \\ \hline
LX / LY / LZ               &  600 / 10 / 2200  & ft        \\ \hline
Initial water saturation   &  0.1              &           \\ \hline
Initial pressure           &  2000             & psi       \\ \hline
Water viscosity            &  1                & cP        \\ \hline
Gas   viscosity            &  0.25             & cP        \\ \hline
Water compressibility      &  0                & 1/psi     \\ \hline
Gas   compressibility      &  6.895e-5         & 1/psi     \\ \hline
Rock  compressibility      &  1e-6             & 1/psi     \\ \hline
Reference pressure         &  0                & psi       \\ \hline
Injection rate             &  37.61            & $\textrm{m}^3/\textrm{d}$  \\ \hline
Production BHP             &  500              & psi          \\ \hline
Time-interval size         &  20               & day    \\ \hline
Total simulation time      &  200              & day    \\ \hline
\end{tabular}
\end{table}

\subsection{Case 5: three-phase, pure viscous, SPE 10 model }

The specification of the three-phase model is shown in Table \ref{tab:specification_WAG}. 

\subsubsection{Case 5a }

\begin{figure}[!htb]
\centering
\subfloat[Oil saturation profile]{
\includegraphics[scale=0.61]{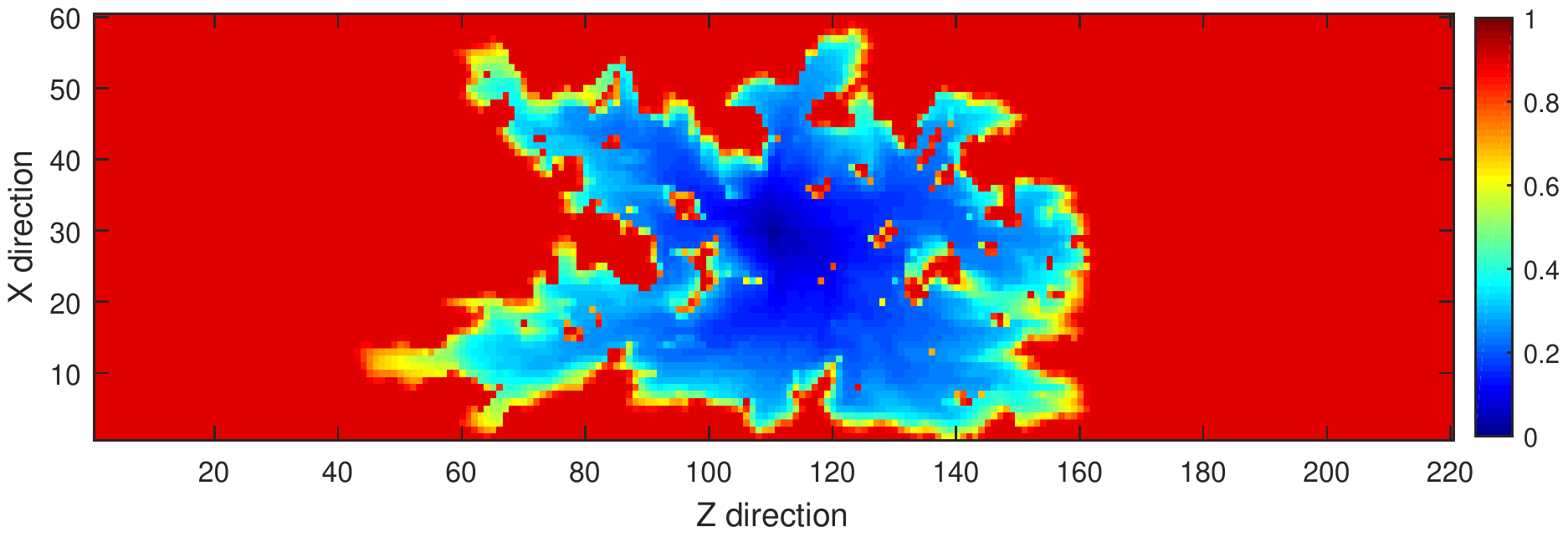}}
\\
\subfloat[Gas saturation profile]{
\includegraphics[scale=0.61]{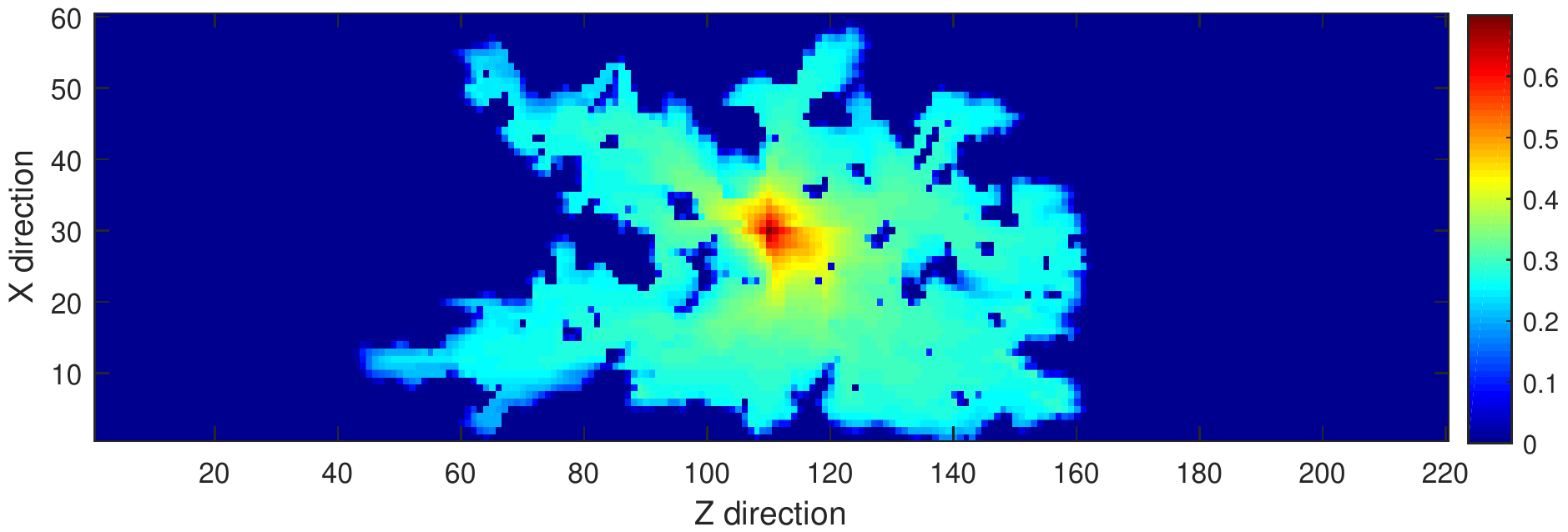}}
\caption{Saturation profiles of Case 5a }
\label{fig:s_WAG_spe}
\end{figure}

\begin{figure}[!htb]
\centering
\includegraphics[scale=0.6]{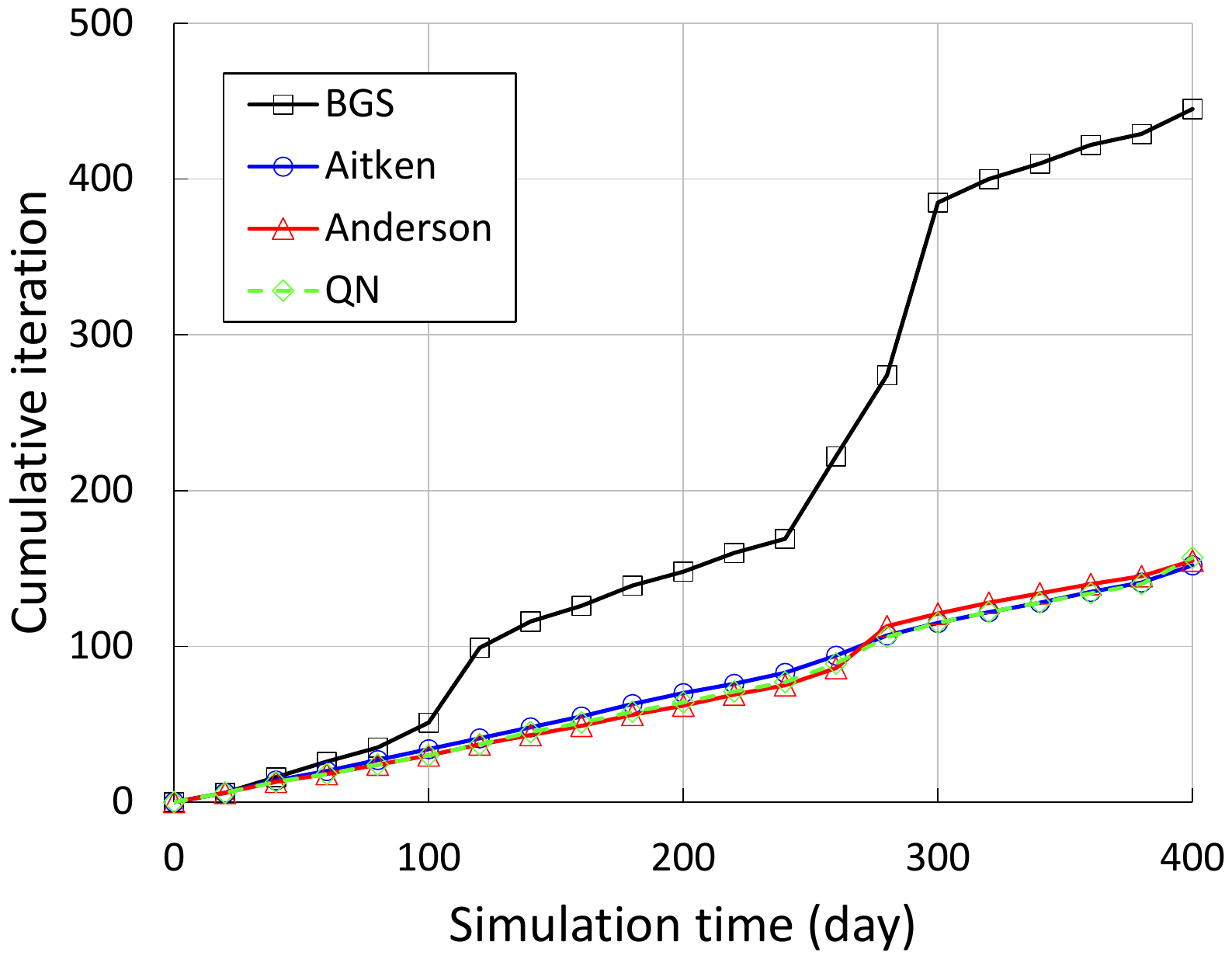}
\caption{Cumulative number of outer iterations versus simulation time of Case 5a }
\label{fig:cN_WAG_spe}
\end{figure}

We first consider a case with water-alternating-gas (WAG) injection. Water is injected first, then followed by gas injection, and the process alternates for each time interval. Cubic relative-permeability functions are used. The oil and gas saturation profiles are plotted in \textbf{Fig. \ref{fig:s_WAG_spe}}. The cumulative number of outer iterations versus simulation time is shown in \textbf{Fig. \ref{fig:cN_WAG_spe}}. The simulation scenario is quite challenging because of the frequently changing well schedule. The nonlinear acceleration methods lead to dramatic improvements upon the BGS process.

\begin{table}[!htb]
\centering
\caption{Specification of the three-phase model}
\label{tab:specification_WAG}
\begin{tabular}{|c|c|c|}
\hline
Parameter                  &  Value            & Unit      \\ \hline
NX / NZ                    &  60 / 220         &           \\ \hline
LX / LY / LZ               &  600 / 10 / 2200  & ft        \\ \hline
Initial water saturation   &  0.1              &           \\ \hline
Initial oil saturation     &  0.9              &           \\ \hline
Initial pressure           &  2000             & psi       \\ \hline
Water viscosity            &  1                & cP        \\ \hline
Oil   viscosity            &  1                & cP        \\ \hline
Gas   viscosity            &  0.25             & cP        \\ \hline
Water compressibility      &  0                & 1/psi     \\ \hline
Oil   compressibility      &  6.895e-6         & 1/psi     \\ \hline
Gas   compressibility      &  6.895e-5         & 1/psi     \\ \hline
Rock  compressibility      &  1e-6             & 1/psi     \\ \hline
Reference pressure         &  0                & psi       \\ \hline
Injection rate             &  37.61            & $\textrm{m}^3/\textrm{d}$  \\ \hline
Production BHP             &  500              & psi          \\ \hline
Time-interval size         &  20               & day    \\ \hline
Total simulation time      &  400              & day    \\ \hline
\end{tabular}
\end{table}

\subsubsection{Case 5b }

\begin{figure}[!htb]
\centering
\includegraphics[scale=0.61]{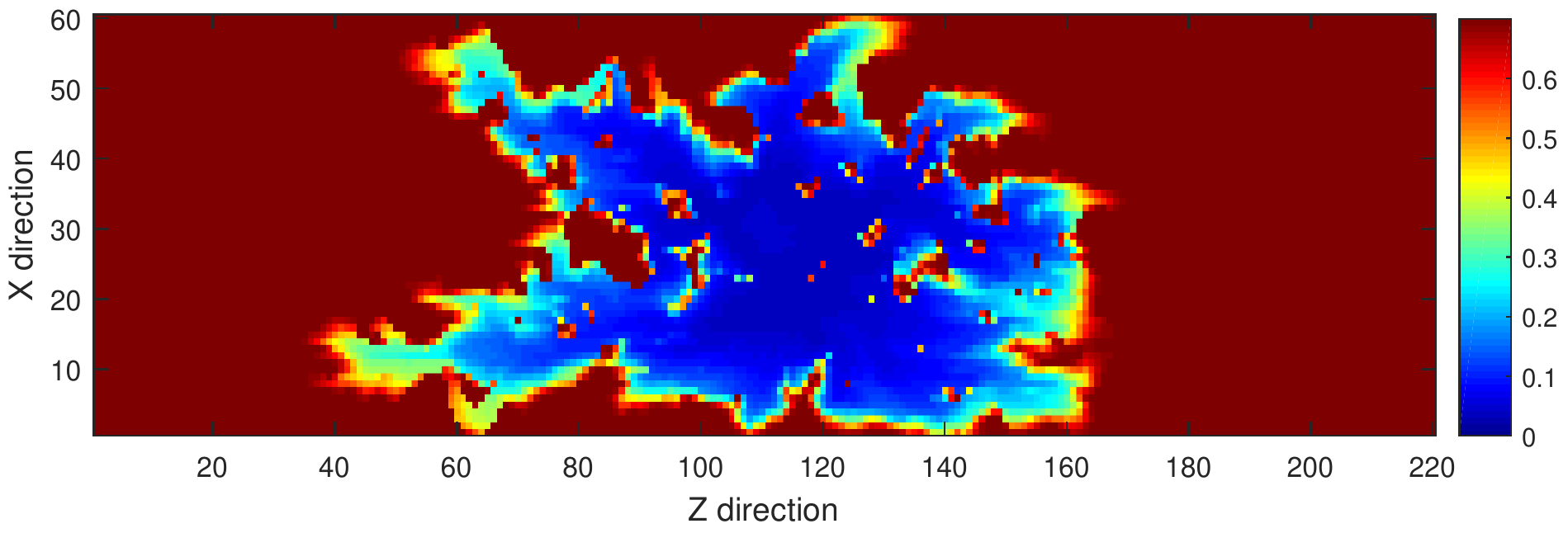}
\caption{Gas saturation profile of Case 5b }
\label{fig:s_3P_2_spe}
\end{figure}

\begin{figure}[!htb]
\centering
\includegraphics[scale=0.6]{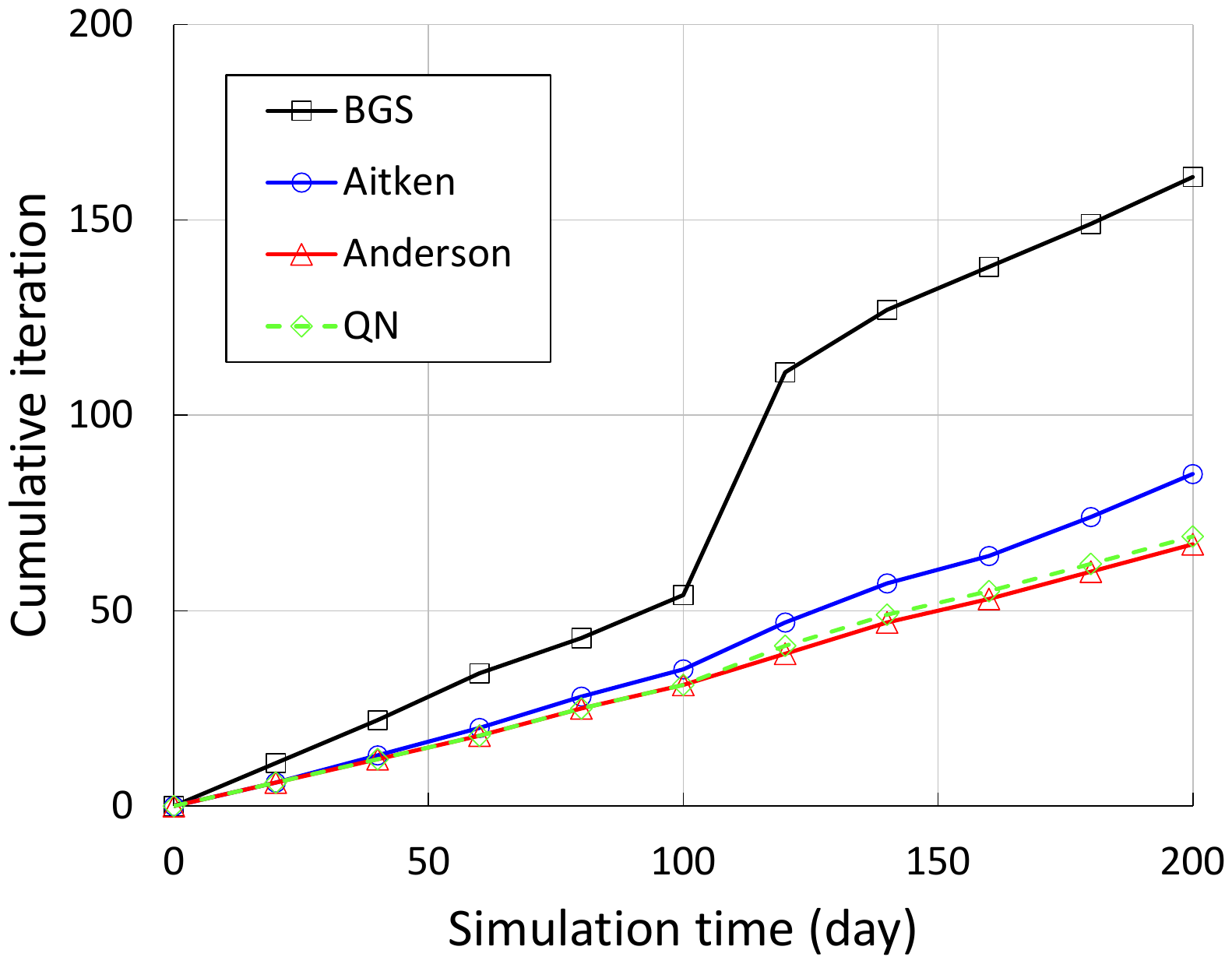}
\caption{Cumulative number of outer iterations versus simulation time of Case 5b }
\label{fig:cN_3P_2_spe}
\end{figure}

We consider a water injection case. The modified model parameters are summarized in Table \ref{tab:specification_WI}. Quadratic relative-permeability functions are used. The three phases are initially present in all the control volumes. The gas saturation profile and the cumulative number of outer iterations versus simulation time are plotted in \textbf{Fig. \ref{fig:s_3P_2_spe}} and \textbf{Fig. \ref{fig:cN_3P_2_spe}}, respectively. BGS undergoes convergence difficulty with several time-step cuts during a short period of simulation. Nevertheless, the sharp rising in the iteration count significantly increases the overall computational cost.

\begin{table}[!htb]
\centering
\caption{Modified model parameters of Case 5b }
\label{tab:specification_WI}
\begin{tabular}{|c|c|c|}
\hline
Parameter                  &  Value            & Unit      \\ \hline
Initial water saturation   &  0.1              &           \\ \hline
Initial oil saturation     &  0.2              &           \\ \hline
Oil   viscosity            &  4                & cP        \\ \hline
Injection rate             &  75.22            & $\textrm{m}^3/\textrm{d}$  \\ \hline
Total simulation time      &  200              & day    \\ \hline
\end{tabular}
\end{table}

\subsubsection{Case 5c }

\begin{figure}[!htb]
\centering
\includegraphics[scale=0.61]{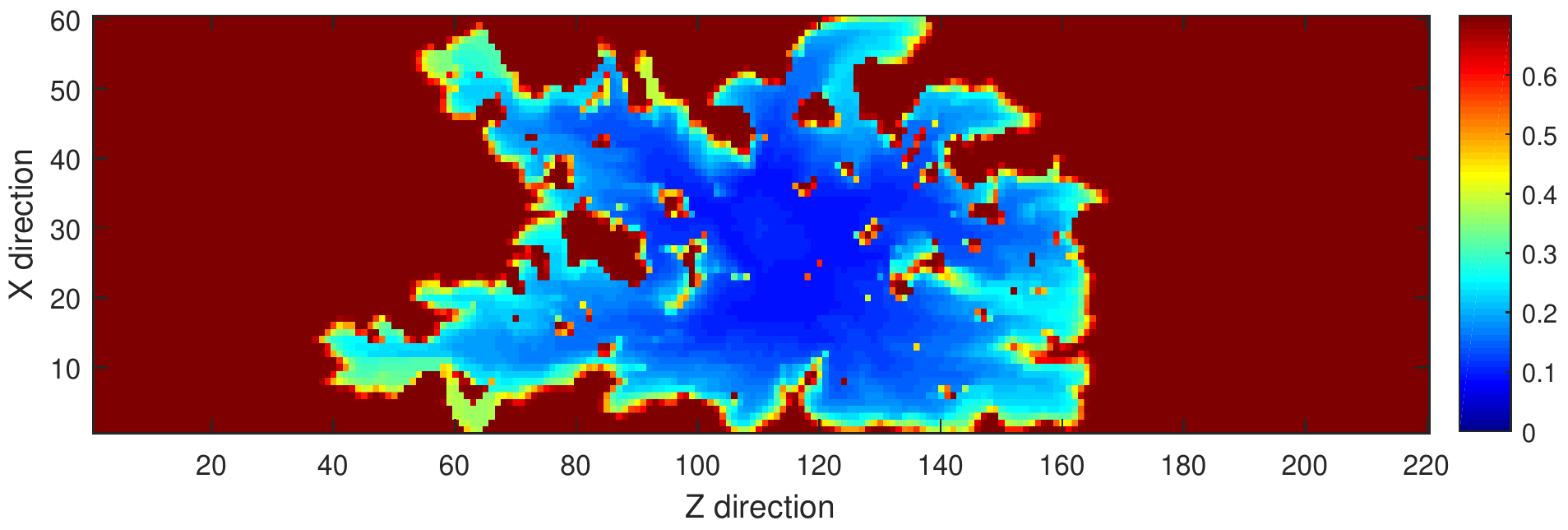}
\caption{Gas saturation profile of Case 5c }
\label{fig:s_3P_3_spe}
\end{figure}

\begin{figure}[!htb]
\centering
\includegraphics[scale=0.6]{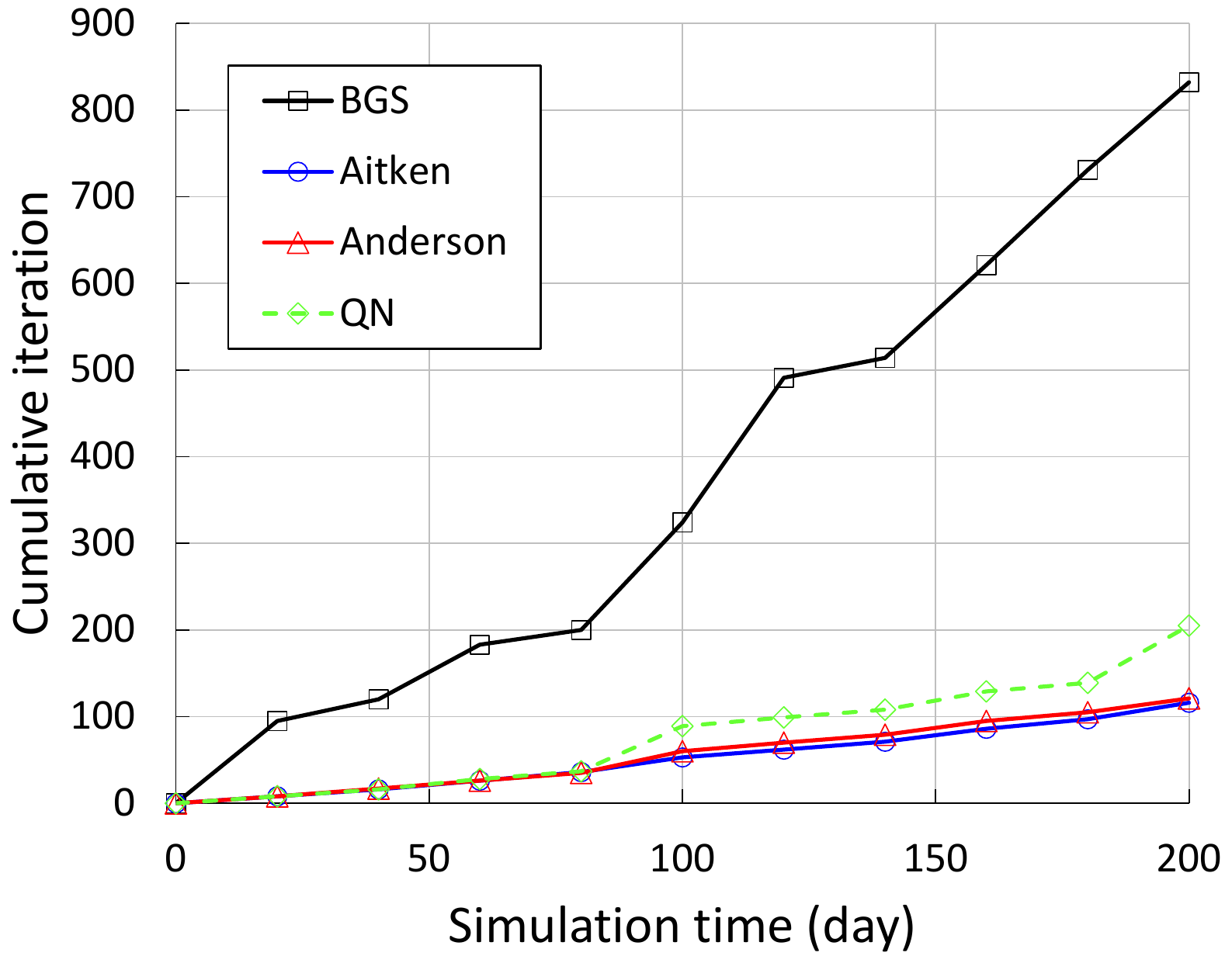}
\caption{Cumulative number of outer iterations versus simulation time of Case 5c }
\label{fig:cN_3P_3_spe}
\end{figure}

Now we test the case with cubic relative-permeability functions. The other model parameters are the same with Case 5b. The gas saturation profile and the cumulative number of outer iterations versus simulation time are plotted in \textbf{Fig. \ref{fig:s_3P_3_spe}} and \textbf{Fig. \ref{fig:cN_3P_3_spe}}, respectively. In the presence of all the three phases, a challenging scenario is generated with strong coupling terms between the flow and transport. As can be seen, the three acceleration methods possess superior convergence properties compared to BGS, which exhibits large numbers of time-step cuts and wasted iterations. Aitken enjoys the optimal number of outer iterations among the studied coupling methods, while QN suffers from a performance deterioration in this simulation case.

\begin{figure}[!htb]
\centering
\includegraphics[scale=0.7]{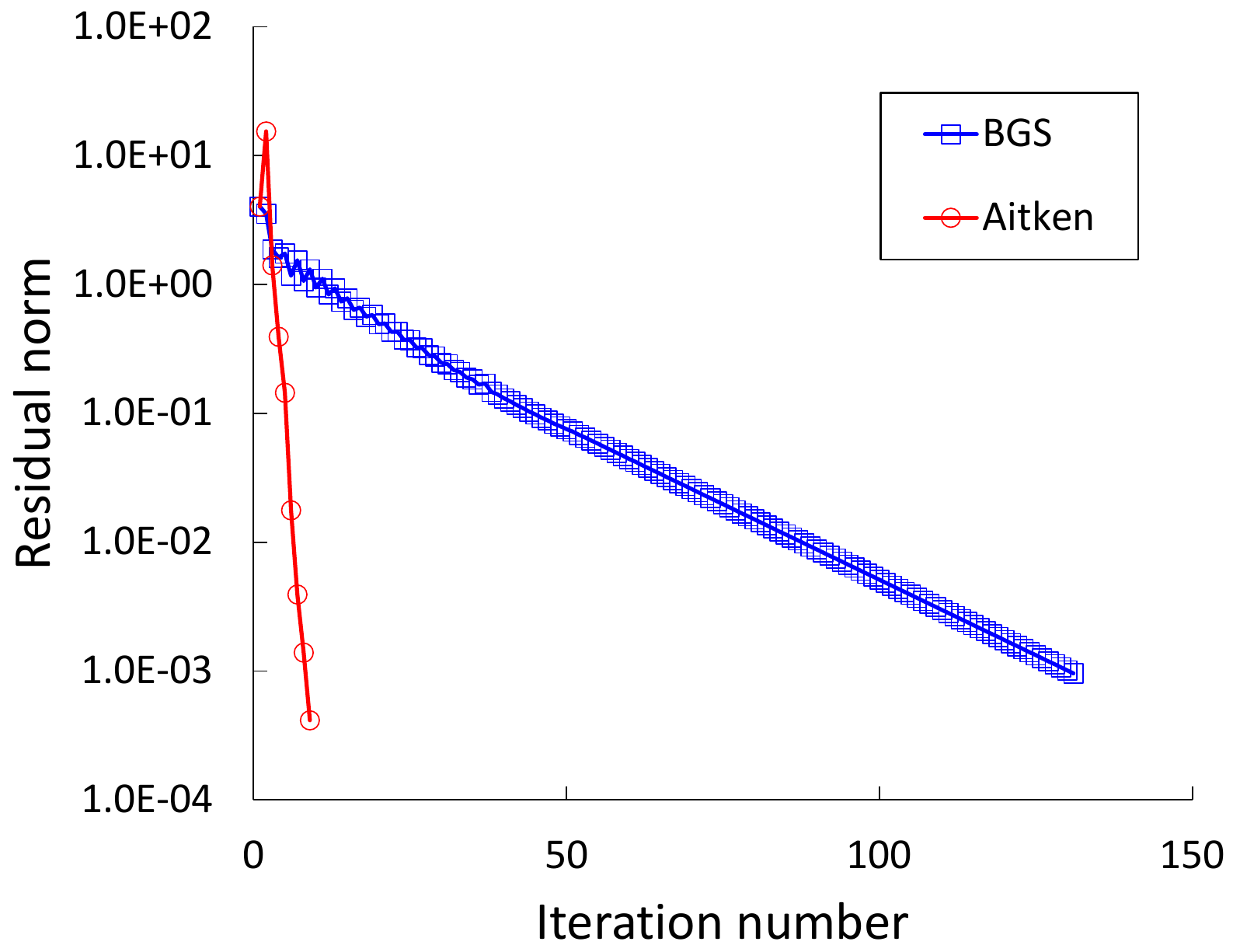}
\caption{Residual history of the time-step size as 20 days for Case 5c }
\label{fig:rh_3P_3_spe}
\end{figure}

We plot the residual history of the time-step size as 20 days in \textbf{Fig. \ref{fig:rh_3P_3_spe}}. A huge contrast in the convergence behavior is observed between Aitken and BGS. Aitken leads to a steep residual decrease, whereas BGS shows a marginal but constant residual decrease between iterations. This behavior represents a typical residual pattern for the problem with strong coupling effects in the phase mobility and total flux terms, compared to the residual stagnation (Case 3b) caused by the flow reversal issue.

\section{Summary}

The sequential fully implicit (SFI) method has been widely employed in the MSFV framework. Each time step for SFI consists of an outer loop to solve the coupled system, in which there is one inner Newton loop to implicitly solve the pressure equation and another loop to implicitly solve the transport equations. Although its implementation is straightforward, the basic SFI method can suffer from severe convergence difficulties, if high coupling degrees exist between the flow and transport sub-problems. In this paper we investigate three nonlinear acceleration techniques to improve the outer-loop convergence, that is, Aitken relaxation, quasi-Newton and Anderson acceleration. The acceleration techniques are adapted and studied for the first time within the context of SFI for coupled flow and transport in porous media. 

We demonstrate the effectiveness of the acceleration techniques using several complex examples. We focus on heterogeneous problems with immiscible multiphase flow and transport. The simulation results show that the performance of SFI degrades significantly for the cases with high coupling degrees. Numerical evidences and detailed analysis are provided to examine the coupling mechanisms between the sub-problems. In the presence of gravitational force, phase upstream direction may keep switching (flow reversal) between the sequential updates. This discontinuous behavior has large impacts on the outer-loop level, leading to oscillations or divergence of the iterative process. We reveal a specific flow reversal issue that is caused by the sign change of the total flux obtained from the pressure equation. Compared with the basic SFI method, superior convergence performance is achieved by the three acceleration techniques, which can resolve the convergence difficulties associated with the various types of coupling effects. We show across a wide range of flow conditions that the studied techniques largely stabilize the iterative process, and reduce the outer iteration count. Among them, Aitken relaxation provides the optimal performance in terms of overall efficiency, because of its low cost for extra computations.

\section*{Acknowledgements}

This work was supported by the Stanford University Petroleum Research Institute for Reservoir Simulation (SUPRI-B).

\section*{Appendix A. Graphical explanations of Aitken's $\Delta^2$ method}

\subsection*{Fixed-point iteration}

\begin{figure}[!htb]
\centering
\includegraphics[scale=0.8]{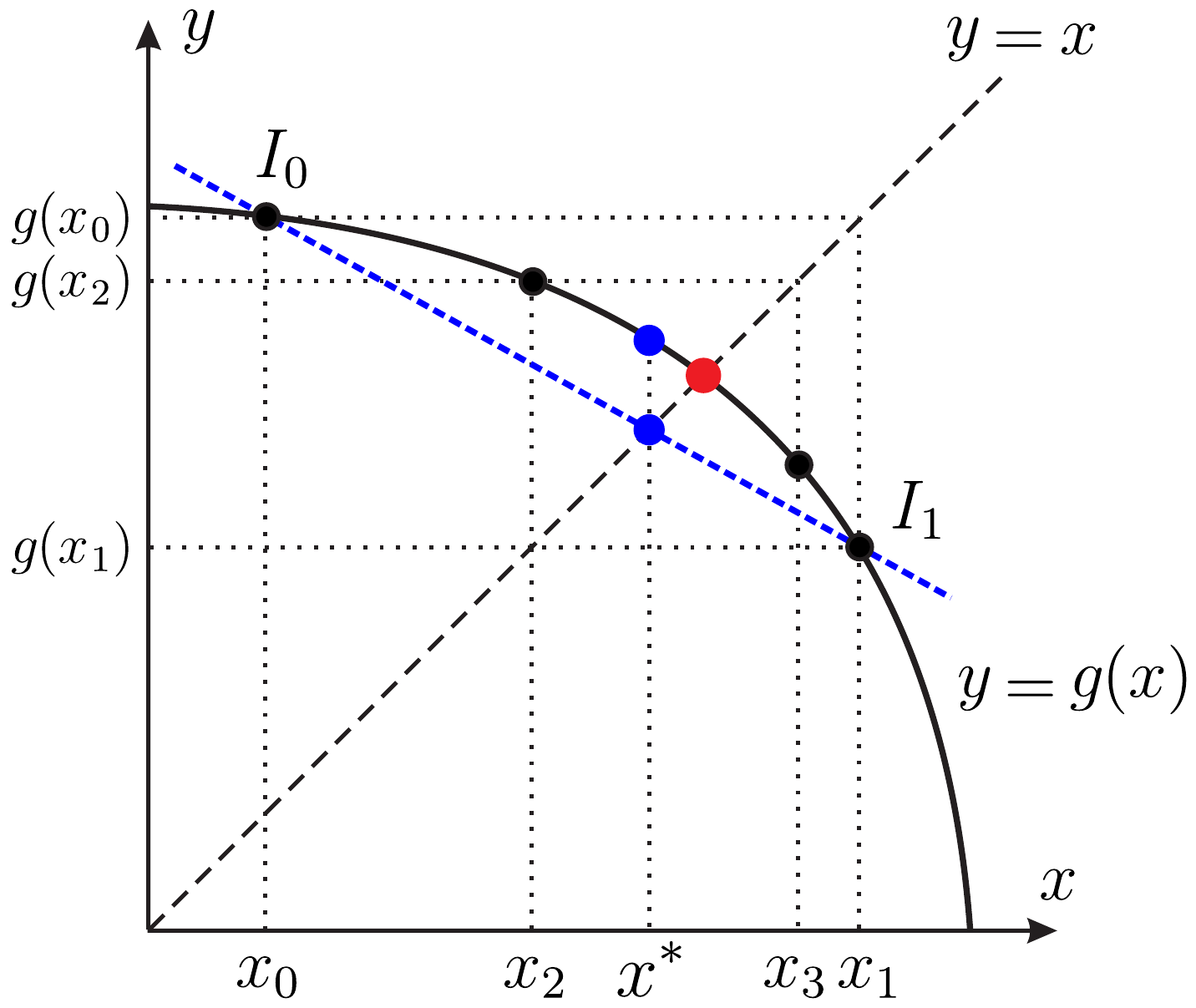}
\caption{Graphical explanation of Aitken's $\Delta^2$ method for fixed-point iteration }
\label{fig:fpi_aitken}
\end{figure}

Suppose we want to solve $x = g(x)$ given $g: \mathbb{R}^n \rightarrow \mathbb{R}^n$. Basic fixed-point iteration (FPI) for this problem is described in Algorithm \ref{alg:FPI}. \textbf{Fig. \ref{fig:fpi_aitken}} shows a graphical illustration of the iterative process. The converged solution of the problem is the intersection point (red dot) of the line $y=x$ with the curve $y = g(x)$. Given an initial value $x_0$, we find the corresponding point on the curve $y = g(x)$, and its $y$ value as $g(x_0)$. Through the convergent sequence $x_{k+1} = g(x_k)$, we now have the value of $x_1$. Then we project $x_1$ onto the $x$ axis with respect to the line $y=x$. Repeating the above procedure we can obtain $x_2$, $x_3$ and so on and so forth. From the figure we see that the sequence gradually approaches the solution, but converges quite slowly.

Now we construct the line going through the first two points $I_0$ and $I_1$. The equation of this line can be expressed as
\begin{equation} 
y = g(x_0) + \left ( x-x_0 \right ) \frac{g(x_1) - g(x_0)}{x_1 - x_0}
\end{equation}
Replacing $g(x_0)$ by $x_1$ and $g(x_1)$ by $x_2$ gives
\begin{equation} 
y = x_1 + \left ( x-x_0 \right ) \frac{x_2 - x_1}{x_1 - x_0}
\end{equation}

In essence, Aitken's $\Delta^2$ method improves the convergence of FPI by providing a better approximation using the intersection of $I_0 I_1$ with $y=x$. The $x$ value $(x^{\ast})$ of the intersection point (blue dot) is obtained by solving 
\begin{equation} 
x = x_1 + \left ( x - x_0 \right ) \frac{x_2 - x_1}{x_1 - x_0}
\end{equation}
Finally we achieve
\begin{equation} 
x = \frac{x_0 x_2 - x_1 x_1}{\left ( x_0 - x_1 \right ) + \left ( x_2 - x_1 \right )}
\end{equation}
which is identical to the scalar formula from Eq. (\ref{eq:Aitken_33_f}) 
\begin{equation} 
s^{(\nu+2)} = \frac{s^{(\nu)}\widetilde{s}^{(\nu+2)} - \widetilde{s}^{(\nu+1)} s^{(\nu+1)}}{s^{(\nu)} - \widetilde{s}^{(\nu+1)} - s^{(\nu+1)} + \widetilde{s}^{(\nu+2)}} 
\end{equation}
if no relaxation is performed so that $\widetilde{s}^{(\nu+1)} = s^{(\nu+1)}$.

Clearly, the approximation $(x^{\ast})$ from Aitken is much closer to the converged solution than the steps generated in the basic FPI process.

\begin{algorithm}
\caption{Fixed-point iteration (FPI)} \label{alg:FPI}
\begin{algorithmic}[1]
\State Given $x_0$
\For{$k=0,1,...$ } 
\State Set $x_{k+1} = g(x_k)$
\EndFor
\end{algorithmic}
\end{algorithm}

\subsection*{Nonlinear block Gauss-Seidel}

Previously we show that the iterative form of SFI is equivalent to the block Gauss-Seidel (BGS) process
\begin{equation} 
\label{eq:new_bgs}
\begin{cases}
p_{k+1} = \mathcal{P} \left ( s_{k}   \right ) \\ 
s_{k+1} = \mathcal{T} \left ( p_{k+1} \right )  
\end{cases}
\end{equation}
where the operators $\mathcal{P}$ and $\mathcal{T}$ represent respectively the pressure and transport solvers. The transport solver works on the results from the pressure solver. Alternate correction is required to bring the pressure and saturation responses into balance so that the global residual vanishes.

\begin{figure}[!htb]
\centering
\includegraphics[scale=0.6]{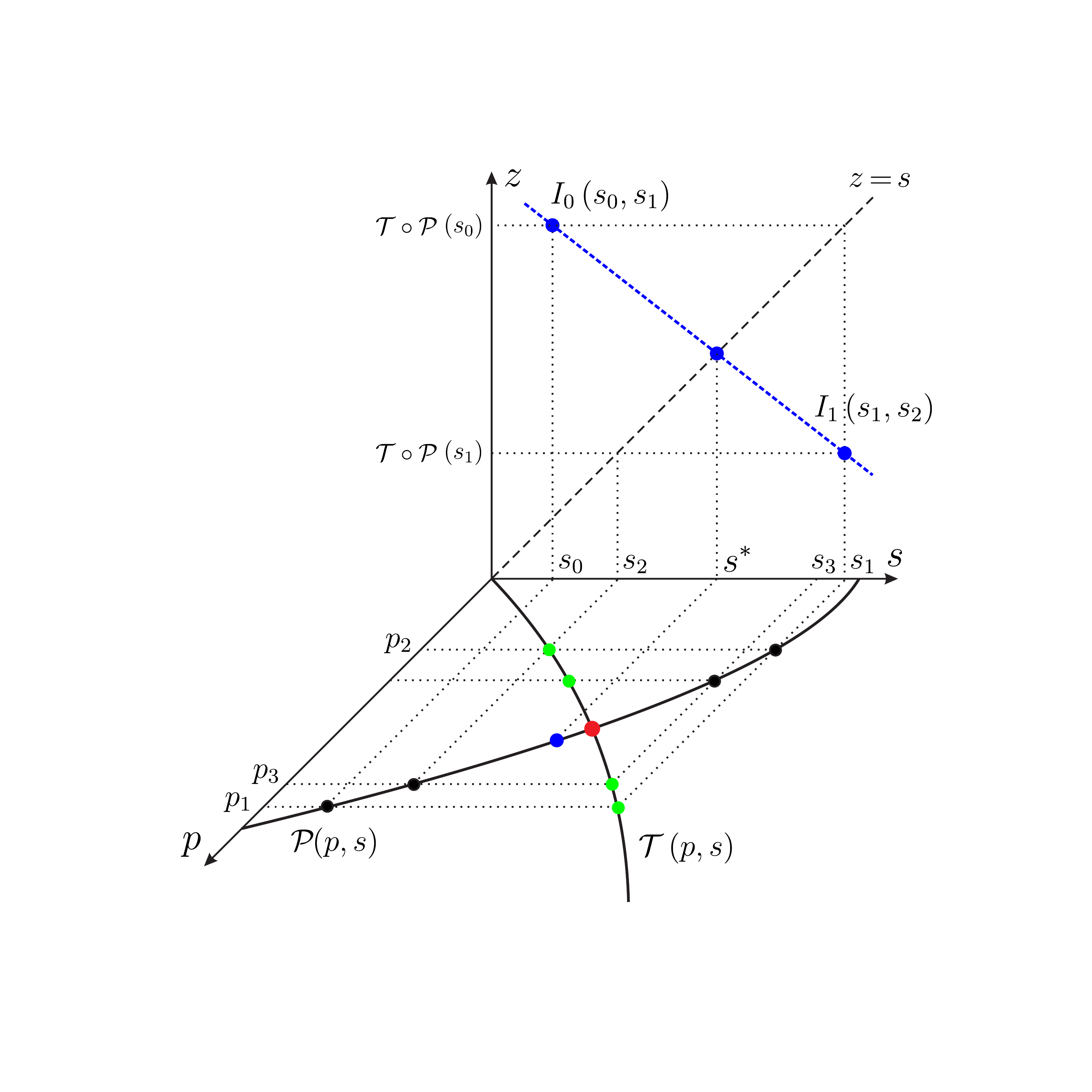}
\caption{Graphical explanation of Aitken's $\Delta^2$ method for the block Gauss-Seidel process }
\label{fig:bgs_aitken}
\end{figure}

We consider a case with two variables ($p$ and $s$). \textbf{Fig. \ref{fig:bgs_aitken}} shows a graphical illustration of the BGS process. The intersection point of the two sub-problems (red dot) represents the converged solution of the global problem. Starting from the initial value $s_0$, we obtain $p_1$ from the curve $\mathcal{P} \left ( p,s \right )$. Then $p_1$ is passed to $\mathcal{T} \left ( p,s \right )$, providing $s_1$. The sequence is generated through repeated applications of the update formulas (\ref{eq:new_bgs}). We can see that the basic BGS iterations approach the solution following a quite convoluted path.

A third dimension has to be added in the figure to present Aitken's $\Delta^2$ method. A BGS iteration can be recast in the FPI form as
\begin{equation} 
s_{k+1} = \mathcal{T} \circ \mathcal{P} \left ( s_{k} \right ) = g(s_{k})
\end{equation}
Again we construct the line going through the first two points $I_0 \left ( s_0, s_1 \right )$ and $I_1 \left ( s_1, s_2 \right )$. A better approximation $(s^{\ast})$ is obtained from the intersection (blue dot) of $I_0 I_1$ with $z=s$. The new point predicted by Aitken is much closer to the point representing the converged solution.

\section*{Appendix B. Quasi-Newton method}

\begin{algorithm}
\caption{Quasi-Newton algorithm} \label{alg:QN_FPI}
\begin{algorithmic}[1]
\State Given $x_0$ and $m \geq 1$
\State Set $x_1 = g(x_0)$
\For{$k=1,2,...$ } 
\State Set $m_k = \textrm{min} \left \{ m,k \right \}$
\State Set $F_k = \left ( f_{k-m_k},...,f_k \right )$, where $f_i = g(x_i) - x_i$
\State Determine $\beta^{k} = \left ( \beta^{k}_0, ..., \beta^{k}_{m_k} \right )^T $ that solves
\State $ \ \quad \textrm{min}_{\beta =\left ( \beta_0,...,\beta_{m_k} \right )^T} \left \| F_k \beta \right \|_2 \quad \textrm{s.t.} \ \sum_{i=0}^{m_k} \beta_i = 1 $ 
\State Set $x_{k+1} = \sum_{i=0}^{m_k} \beta_{i}^{k} g(x_{k-m_k+i}) $
\EndFor
\end{algorithmic}
\end{algorithm}

The fixed-point iterative method can achieve nonlinear solutions without requiring any derivative information. However, the convergence rate of FPI generally is very low. Quasi-Newton (QN) method is a popular class of nonlinear acceleration techniques to improve the convergence of FPI. A general form of QN is summarized in Algorithm \ref{alg:QN_FPI}. The constrained linear least-squares problem in Algorithm \ref{alg:QN_FPI} can be recast in an unconstrained form, which is straightforward to solve (Walker and Ni 2011). We define $\Delta f_i = f_{i+1} - f_i $ for each $i$ and set $F_k = \left ( \Delta f_{k-m_k}, ..., \Delta f_{k-1} \right ) $. Then the least-squares problem is equivalent to
\begin{equation} 
\label{eq:LS_p}
\underset{\gamma=\left ( \gamma_0, ..., \gamma_{m_k-1} \right )^T}{\textrm{min}} \left \| f_k - F_k \gamma \right \|_2 
\end{equation}
where $\beta $ and $\gamma $ are related by $\beta_0 = \gamma_0 $, $\beta_i = \gamma_i - \gamma_{i-1}$ for $1 \leq i \leq m_k - 1$, and $\beta_{m_k} = 1 - \gamma_{m_k-1}$. With the solution of (\ref{eq:LS_p}) denoted by $\gamma^{k} =\left ( \gamma_0^{k},...,\gamma_{m_k-1}^{k} \right )^T $, the next iterate of QN is written as
\begin{equation} 
\label{eq:QN_iter}
x_{k+1} = x_{k} + f_k - \left ( X_k + F_k \right ) \gamma^{k} 
\end{equation} 
where $X_k = \left ( \Delta x_{k-m_k}, ..., \Delta x_{k-1} \right )$ with $\Delta x_i = x_{i+1} - x_i$ for each $i$. Substituting the normal-equation characterization $\gamma^k = \left ( X_{k}^{T} F_k \right )^{-1} X_{k}^{T} f_k $ in expression (\ref{eq:QN_iter}), one obtains the quasi-Newton form (Walker and Ni 2011)
\begin{equation} 
x_{k+1} = x_k - \widehat{J_{k}}^{-1} f_k 
\end{equation} 
where
\begin{equation} 
\widehat{J_{k}}^{-1} = -I + \left ( X_k + F_k \right ) \left ( X_{k}^{T} F_k \right )^{-1} X_{k}^{T} 
\end{equation} 
is regarded as an approximate inverse of the Jacobian of $f(x) = g(x) - x$. The approximate Jacobian satisfies the direct multisecant condition $\widehat{J_{k}} X_k = F_k$. $\widehat{J_{k}}$ is the $\textit{first Broyden update}$ of $-I$ satisfying $\widehat{J_{k}} X_k = F_k$ (Fang and Saad 2009).

\section*{References}

Aziz, K., Settari, A., 1979. Petroleum Reservoir Simulation. Chapman \& Hall.

An, H., Jia, X. and Walker, H.F., 2017. Anderson acceleration and application to the three-temperature energy equations. Journal of Computational Physics, 347, pp.1-19.

Baker, L.E., 1988, January. Three-phase relative permeability correlations. In SPE Enhanced Oil Recovery Symposium. Society of Petroleum Engineers.

Coats, K.H., 2000. A note on IMPES and some IMPES-based simulation models. SPE Journal, 5(03), pp.245-251.

Cao, H., 2002. Development of techniques for general purpose simulators (Doctoral dissertation, Stanford University).

Cusini, M., Lukyanov, A.A., Natvig, J. and Hajibeygi, H., 2015. Constrained pressure residual multiscale (CPR-MS) method for fully implicit simulation of multiphase flow in porous media. Journal of Computational Physics, 299, pp.472-486.

Degroote, J., Souto-Iglesias, A., Van Paepegem, W., Annerel, S., Bruggeman, P. and Vierendeels, J., 2010. Partitioned simulation of the interaction between an elastic structure and free surface flow. Computer methods in applied mechanics and engineering, 199(33), pp.2085-2098.

Efendiev, Y., Ginting, V., Hou, T. and Ewing, R., 2006. Accurate multiscale finite element methods for two-phase flow simulations. Journal of Computational Physics, 220(1), pp.155-174.

Efendiev, Y., Galvis, J. and Hou, T.Y., 2013. Generalized multiscale finite element methods (GMsFEM). Journal of Computational Physics, 251, pp.116-135.

Erbts, P. and DuSter, A., 2012. Accelerated staggered coupling schemes for problems of thermoelasticity at finite strains. Computers $\&$ Mathematics with Applications, 64(8), pp.2408-2430.

Fang, H.R. and Saad, Y., 2009. Two classes of multisecant methods for nonlinear acceleration. Numerical Linear Algebra with Applications, 16(3), pp.197-221.

Helmig, R., Niessner, J., Flemisch, B., Wolff, M. and Fritz, J., 2010. Efficient Modeling of Flow and Transport in Porous Media Using Multiphysics and Multiscale Approaches. In Handbook of Geomathematics (pp. 417-457). Springer Berlin Heidelberg.

Hamon, F.P. and Tchelepi, H.A., 2016. Analysis of hybrid upwinding for fully-implicit simulation of three-phase flow with gravity. SIAM Journal on Numerical Analysis, 54(3), pp.1682-1712.

Hamon, F.P., Mallison, B.T. and Tchelepi, H.A., 2016. Implicit Hybrid Upwind scheme for coupled multiphase flow and transport with buoyancy. Computer Methods in Applied Mechanics and Engineering, 311, pp.599-624.

Irons B and Tuck RC, 1969. A version of the Aitken accelerator for computer implementation. Int J Numer Methods Eng 1:275–277

Jenny, P., Lee, S.H. and Tchelepi, H.A., 2003. Multi-scale finite-volume method for elliptic problems in subsurface flow simulation. Journal of Computational Physics, 187(1), pp.47-67.

Jenny, P., Lee, S.H. and Tchelepi, H.A., 2006. Adaptive fully implicit multi-scale finite-volume method for multi-phase flow and transport in heterogeneous porous media. Journal of Computational Physics, 217(2), pp.627-641.

Jenny, P., Tchelepi, H.A. and Lee, S.H., 2009. Unconditionally convergent nonlinear solver for hyperbolic conservation laws with S-shaped flux functions. Journal of Computational Physics, 228(20), pp.7497-7512.

Jeannin, L., Mainguy, M., Masson, R. and Vidal-Gilbert, S., 2007. Accelerating the convergence of coupled geomechanical‐reservoir simulations. International journal for numerical and analytical methods in geomechanics, 31(10), pp.1163-1181. 

Jiang, J. and Younis, R.M., 2017. Efficient C1-continuous phase-potential upwind (C1-PPU) schemes for coupled multiphase flow and transport with gravity. Advances in Water Resources, 108, pp.184-204.

Kuttler, U. and Wall, W.A., 2008. Fixed-point fluid–structure interaction solvers with dynamic relaxation. Computational Mechanics, 43(1), pp.61-72.

Krogstad, S., Lie, K.A., Nilsen, H.M., Natvig, J.R., Skaflestad, B. and Aarnes, J.E., 2009, January. A multiscale mixed finite element solver for three phase black oil flow. In SPE Reservoir Simulation Symposium. Society of Petroleum Engineers.

Kozlova, A., Li, Z., Natvig, J.R., Watanabe, S., Zhou, Y., Bratvedt, K. and Lee, S.H., 2016. A real-field multiscale black-oil reservoir simulator. SPE Journal.

Li, B., Chen, Z. and Huan, G., 2003. The sequential method for the black-oil reservoir simulation on unstructured grids. Journal of Computational physics, 192(1), pp.36-72.

Lunati, I. and Jenny, P., 2006. Multiscale finite-volume method for compressible multiphase flow in porous media. Journal of Computational Physics, 216(2), pp.616-636.

Lunati, I. and Jenny, P., 2008. Multiscale finite-volume method for density-driven flow in porous media. Computational Geosciences, 12(3), pp.337-350.

Lu, B., 2008. Iteratively coupled reservoir simulation for multiphase flow in porous media. The University of Texas at Austin.

Lee, S.H., Wolfsteiner, C. and Tchelepi, H.A., 2008. Multiscale finite-volume formulation for multiphase flow in porous media: black oil formulation of compressible, three-phase flow with gravity. Computational Geosciences, 12(3), pp.351-366.

Li, B. and Tchelepi, H.A., 2015. Nonlinear analysis of multiphase transport in porous media in the presence of viscous, buoyancy, and capillary forces. Journal of Computational Physics, 297, pp.104-131.

Lee, S.H., Efendiev, Y. and Tchelepi, H.A., 2015. Hybrid upwind discretization of nonlinear two-phase flow with gravity. Advances in Water Resources, 82, pp.27-38.

Lee, S.H. and Efendiev, Y., 2016. C1-Continuous relative permeability and hybrid upwind discretization of three phase flow in porous media. Advances in Water Resources, 96, pp.209-224.

Lie, K.A., M{\o}yner, O., Natvig, J.R., Kozlova, A., Bratvedt, K., Watanabe, S. and Li, Z., 2017. Successful application of multiscale methods in a real reservoir simulator environment. Computational Geosciences, 21(5-6), pp.981-998.

M{\o}yner, O. and Lie, K.A., 2014. A multiscale two-point flux-approximation method. Journal of Computational Physics, 275, pp.273-293.

M{\o}yner, O. and Lie, K.A., 2016. A multiscale restriction-smoothed basis method for compressible black-oil models. SPE Journal, 21(06), pp.2-079.

Moncorgé, A., Tchelepi, H.A. and Jenny, P., 2017. Modified sequential fully implicit scheme for compositional flow simulation. Journal of Computational Physics, 337, pp.98-115.

Stone, H.L., 1973. Estimation of three-phase relative permeability and residual oil data. J. Pet. Technol.;(United States), 12(4).

Schlumberger: ECLIPSE 2013.2 Technical Description (2013)

Watts, J.W., 1986. A compositional formulation of the pressure and saturation equations. SPE (Society of Petroleum Engineers) Reserv. Eng.;(United States), 1(3).

Walker, H.F. and Ni, P., 2011. Anderson acceleration for fixed-point iterations. SIAM Journal on Numerical Analysis, 49(4), pp.1715-1735.

Wang, X. and Tchelepi, H.A., 2013. Trust-region based solver for nonlinear transport in heterogeneous porous media. Journal of Computational Physics, 253, pp.114-137.

Watanabe, S., Li, Z., Bratvedt, K., Lee, S.H. and Natvig, J., 2016, August. A Stable Multi-phase Nonlinear Transport Solver with Hybrid Upwind Discretization in Multiscale Reservoir Simulator. In ECMOR XV-15th European Conference on the Mathematics of Oil Recovery.

Younis, R., Tchelepi, H.A. and Aziz, K., 2010. Adaptively Localized Continuation-Newton Method--Nonlinear Solvers That Converge All the Time. SPE Journal, 15(02), pp.526-544.

\end{document}